\documentclass[12pt,a4paper]{article}

\usepackage{lineno}  

\usepackage{graphicx}  

\usepackage{xspace}
\usepackage{color}
\usepackage{colortbl}
\usepackage{amsmath,amssymb}

\usepackage{ifthen} 

\newboolean{pdflatex}
\setboolean{pdflatex}{false} 
\newboolean{uprightparticles}
\setboolean{uprightparticles}{false} 
\usepackage{amssymb}
\usepackage{amsfonts}
\usepackage{upgreek}
\def\dzdzb{{~\raise.85em\hbox{{\tiny{(}\textemdash\tiny{)}}}\kern-1.05em
          \lower0.0em\hbox{$D^0$}~}\xspace}

\def\ux85 {UX85\xspace}

\ifthenelse{\boolean{uprightparticles}}%
{

 \def\PDelta      {\ensuremath{\Delta}\xspace}                 
 \def\PXi      {\ensuremath{\Xi}\xspace}                 
 \def\PLambda      {\ensuremath{\Lambda}\xspace}                 
 \def\PSigma      {\ensuremath{\Sigma}\xspace}                 
 \def\POmega      {\ensuremath{\Omega}\xspace}                 
 \def\PUpsilon      {\ensuremath{\Upsilon}\xspace}                 
 
 \def\PB      {\ensuremath{\mathrm{B}}\xspace}                 
                  
 \def\PD      {\ensuremath{\mathrm{D}}\xspace}

 \def\PK      {\ensuremath{\mathrm{K}}\xspace}

 \def\Pc      {\ensuremath{\mathrm{c}}\xspace}

 \def\Pi      {\ensuremath{\mathrm{i}}\xspace}

}
{

 \mathchardef\PDelta="7101
 \mathchardef\PXi="7104
 \mathchardef\PLambda="7103
 \mathchardef\PSigma="7106
 \mathchardef\POmega="710A
 \mathchardef\PUpsilon="7107
                  
 \def\PB      {\ensuremath{B}\xspace}                 
                  
 \def\PD      {\ensuremath{D}\xspace}

 \def\PK      {\ensuremath{K}\xspace}

 \def\Pc      {\ensuremath{c}\xspace}

 \def\Pi      {\ensuremath{i}\xspace}

}

\def\c     {\ensuremath{\Pc}\xspace}

\def\kaon  {\ensuremath{\PK}\xspace}
  \def\Kbar  {\kern 0.2em\overline{\kern -0.2em \PK}{}\xspace}

\def\Kz    {\ensuremath{\kaon^0}\xspace}
\def\Kzb   {\ensuremath{\Kbar^0}\xspace}
\def\KzKzb {\ensuremath{\Kz \kern -0.16em \Kzb}\xspace}
\def\Kp    {\ensuremath{\kaon^+}\xspace}
\def\Km    {\ensuremath{\kaon^-}\xspace}

\def\KpKm  {\ensuremath{\Kp \kern -0.16em \Km}\xspace}

  \def\Dbar    {\kern 0.2em\overline{\kern -0.2em \PD}{}\xspace}
\def\D       {\ensuremath{\PD}\xspace}

\def\Dz      {\ensuremath{\D^0}\xspace}
\def\Dzb     {\ensuremath{\Dbar^0}\xspace}
\def\DzDzb   {\ensuremath{\Dz {\kern -0.16em \Dzb}}\xspace}
\def\Dp      {\ensuremath{\D^+}\xspace}
\def\Dm      {\ensuremath{\D^-}\xspace}

\def\DpDm    {\ensuremath{\Dp {\kern -0.16em \Dm}}\xspace}

\def\B       {\ensuremath{\PB}\xspace}
  \def\Bbar    {\kern 0.18em\overline{\kern -0.18em \PB}{}\xspace}

\def\Bz      {\ensuremath{\B^0}\xspace}
\def\Bzb     {\ensuremath{\Bbar^0}\xspace}

\def\Bsb     {\ensuremath{\Bbar^0_s}\xspace}

  \def\Y#1S{\ensuremath{\PUpsilon{(#1S)}}\xspace}

\def\Lb      {\ensuremath{\L_b}\xspace}

\def\Lc      {\ensuremath{\L_c}\xspace}

\def\to                 {\ensuremath{\rightarrow}\xspace}

\def\AT#1     {\ensuremath{A_T^{#1}}\xspace}           

\def\C#1      {\ensuremath{\mathcal{C}_{#1}}\xspace}                       
\def\Cp#1     {\ensuremath{\mathcal{C}_{#1}^{'}}\xspace}                    
\def\Ceff#1   {\ensuremath{\mathcal{C}_{#1}^{\mathrm{(eff)}}}\xspace}        
\def\Cpeff#1  {\ensuremath{\mathcal{C}_{#1}^{'\mathrm{(eff)}}}\xspace}       
\def\Ope#1    {\ensuremath{\mathcal{O}_{#1}}\xspace}                       
\def\Opep#1   {\ensuremath{\mathcal{O}_{#1}^{'}}\xspace}                    

\newcommand{\tev}{\ensuremath{\mathrm{\,Te\kern -0.1em V}}\xspace}
\newcommand{\gev}{\ensuremath{\mathrm{\,Ge\kern -0.1em V}}\xspace}
\newcommand{\mev}{\ensuremath{\mathrm{\,Me\kern -0.1em V}}\xspace}
\newcommand{\kev}{\ensuremath{\mathrm{\,ke\kern -0.1em V}}\xspace}
\newcommand{\ev}{\ensuremath{\mathrm{\,e\kern -0.1em V}}\xspace}
\newcommand{\gevc}{\ensuremath{{\mathrm{\,Ge\kern -0.1em V\!/}c}}\xspace}
\newcommand{\mevc}{\ensuremath{{\mathrm{\,Me\kern -0.1em V\!/}c}}\xspace}
\newcommand{\gevcc}{\ensuremath{{\mathrm{\,Ge\kern -0.1em V\!/}c^2}}\xspace}
\newcommand{\gevgevcccc}{\ensuremath{{\mathrm{\,Ge\kern -0.1em V^2\!/}c^4}}\xspace}
\newcommand{\mevcc}{\ensuremath{{\mathrm{\,Me\kern -0.1em V\!/}c^2}}\xspace}

\def\gsim{{~\raise.15em\hbox{$>$}\kern-.85em
          \lower.35em\hbox{$\sim$}~}\xspace}
\def\lsim{{~\raise.15em\hbox{$<$}\kern-.85em
          \lower.35em\hbox{$\sim$}~}\xspace}

\def\tell1  {TELL1\xspace}
\def\ukl1   {UKL1\xspace}

\def\dzdzb{{~\raise.85em\hbox{{\tiny{(}\textemdash\tiny{)}}}\kern-1.05em
          \lower0.0em\hbox{$D^0$}~}\xspace}
\def\bsbsb{{~\raise.85em\hbox{{\tiny{(}\textemdash\tiny{)}}}\kern-1.05em
          \lower0.0em\hbox{$B_s^0$}~}\xspace}

\def\bzb{\Bzb}

\def\br{{\cal{B}}}

\def\xb{H_b}
\def\xc{H_c}
\def\Lb{\Lambda_b^0}
\def\Lc{\Lambda_c^+}

\def\Sc{\Sigma_c}

\def\xbtoxcpipipi{\xb\to\xc\pi^-\pi^+\pi^-}
\def\xbtoxcpi{\xb\to \xc\pi^-}

\def\bstodspipipi{\Bsb\to D_s^+\pi^-\pi^+\pi^-}
\def\bstodskpipi{\Bsb\to D_s^{\pm}K^{\mp}\pi^{\pm}\pi^{\mp}}
\def\btodpipipi{\Bzb\to D^+\pi^-\pi^+\pi^-}

\def\btodzeropipipi{B^-\to D^0\pi^-\pi^+\pi^-}
\def\btodzerokpipi{B^-\to D K^-\pi^+\pi^-}
\def\btodstarpipipi{\Bzb\to D^{*+}\pi^-\pi^+\pi^-}
\def\LbtoLcpipipi{\Lambda_b^0\to\Lambda_c^+\pi^-\pi^+\pi^-}

\def\LbtoLcpi{\Lambda_b^0\to\Lambda_c^+\pi^-}

\def\bstodspi{\Bsb\to D_s^+\pi^-}
\def\bstodsk{\Bsb\to D_s^{\pm}K^{\mp}}
\def\btodpi{\Bzb\to D^+\pi^-}

\def\btodzeropi{B^{-}\to D^0\pi^{-}}
\def\btodzerok{B^-\to D K^-}

\def\eff{\epsilon}
\def\btodzerokstar{\Bzb\to D K^{*0}}

\def\ipb{\rm pb^{-1}}

\usepackage{hyperref}

\begin{document}



\begin{titlepage}
\pagenumbering{roman}

\vspace*{-1.5cm}
\centerline{\large EUROPEAN ORGANIZATION FOR NUCLEAR RESEARCH (CERN)}
\vspace*{0.5cm}
\hspace*{-0.5cm}
\begin{tabular*}{\linewidth}{lc@{\extracolsep{\fill}}r}
\ifthenelse{\boolean{pdflatex}}
{\vspace*{-2.7cm}\mbox{\!\!\!\includegraphics[width=.14\textwidth]{./lhcb-logo.pdf}} & &}%
{\vspace*{-1.2cm}\mbox{\!\!\!\includegraphics[width=.12\textwidth]{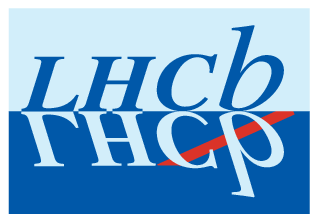}} & &}%
\\
 & & LHCb-PAPER-2011-016 \\  
 & & CERN-PH-EP-2011-151 \\
 & & \today \\ 
 & & \\
\end{tabular*}

\vspace*{1.0cm}

{\bf\boldmath\huge
\begin{center}
  Measurements of the Branching fractions for $B_{(s)}\to D_{(s)}\pi\pi\pi$
and $\Lambda_b^0\to\Lambda_c^+\pi\pi\pi$
\end{center}
}

\vspace*{0.2cm}

\begin{center}
The LHCb Collaboration
\footnote{Authors are listed on the following pages.}
\end{center}


\begin{abstract}
  \noindent

\noindent Branching fractions of the decays $\xbtoxcpipipi$
relative to $\xbtoxcpi$ are presented, where $\xb$ ($\xc$) represents $\bar{B}^0$ ($D^+$), $B^-$ ($D^0$), $\bar{B}_s^0$ ($D_s^+$) 
and $\Lambda_b^0$ ($\Lambda_c^+$). The measurements are performed with the LHCb detector using 
35~$\ipb$ of data collected at $\sqrt{s}=7$~TeV. The ratios of branching fractions are measured to be

\begin{align*}
{\br(\btodpipipi)\over\br(\btodpi)} = 2.38\pm0.11\pm0.21 \nonumber \\
{\br(\btodzeropipipi)\over\br(\btodzeropi)} = 1.27\pm0.06\pm0.11 \nonumber \\
{\br(\bstodspipipi)\over\br(\bstodspi)} = 2.01\pm0.37\pm0.20 \nonumber \\
{\br(\LbtoLcpipipi)\over\br(\LbtoLcpi)} = 1.43\pm0.16\pm0.13. \nonumber \\
\end{align*}

\noindent We also report measurements of partial decay rates of these decays to excited charm hadrons.
These results are of comparable or higher precision than existing measurements.
\end{abstract}


\end{titlepage}



{\footnotesize{
\noindent 
R.~Aaij$^{23}$, 
B.~Adeva$^{36}$, 
M.~Adinolfi$^{42}$, 
C.~Adrover$^{6}$, 
A.~Affolder$^{48}$, 
Z.~Ajaltouni$^{5}$, 
J.~Albrecht$^{37}$, 
F.~Alessio$^{37}$, 
M.~Alexander$^{47}$, 
G.~Alkhazov$^{29}$
% ...

{\it\noindent
% ...
$ ^{5}$Clermont Universit\'{e}, Universit\'{e} Blaise Pascal, CNRS/IN2P3, LPC, Clermont-Ferrand, France\\
$ ^{6}$CPPM, Aix-Marseille Universit\'{e}, CNRS/IN2P3, Marseille, France\\
% ...
$ ^{23}$Nikhef National Institute for Subatomic Physics, Amsterdam, Netherlands\\
% ...
$ ^{29}$Petersburg Nuclear Physics Institute (PNPI), Gatchina, Russia\\
% ...
$ ^{36}$Universidad de Santiago de Compostela, Santiago de Compostela, Spain\\
$ ^{37}$European Organization for Nuclear Research (CERN), Geneva, Switzerland\\
% ...
$ ^{42}$H.H. Wills Physics Laboratory, University of Bristol, Bristol, United Kingdom\\
% ...
$ ^{47}$School of Physics and Astronomy, University of Glasgow, Glasgow, United Kingdom\\
$ ^{48}$Oliver Lodge Laboratory, University of Liverpool, Liverpool, United Kingdom\\
% ...
}}}
\bigskip

\cleardoublepage




\pagestyle{plain} 
\setcounter{page}{1}
\pagenumbering{arabic}


%



\section{Introduction}

Over the last two decades, a wealth of information has been accumulated on the decays of $b$-hadrons.
Measurements of their decays have been used to test the CKM mechanism~\cite{ckm} for describing
weak decay phenomena in the Standard Model, as well as provide measurements against which
various theoretical approaches, such as HQET~\cite{hqet} and the factorization hypothesis,
can be compared.
While many decays have been measured, a large number remain either unobserved or poorly measured,
most notably in the decays of $B^0_s$ mesons and $\Lambda_b^0$ baryons. Among the largest
hadronic branching fractions are the decays $\xbtoxcpipipi$, where 
$\xb$ ($\xc$) represents $\Bzb$ ($D^+$), $B^-$ ($D^0$), $\Bsb$ ($D_s^+$) 
and $\Lambda_b^0$ ($\Lambda_c^+$). The first three branching fractions were determined with 
only 30-40\% accuracy, and the $\LbtoLcpipipi$ branching fraction was unmeasured.

Beyond improving our overall understanding of hadronic $b$ decays, these decays are of 
interest because of their potential use in {\it CP} violation studies.
It is well known that the Cabibbo-suppressed decays $\btodzerok$~\cite{ads,glw,ggsz} and $\bstodsk$~\cite{dsk1,dsk2} 
provide clean measurements of the weak phase $\gamma$ through time-independent and time-dependent
rate measurements, respectively. Additional sensitivity can be obtained by using $\btodpi$~\cite{b2dpicp} decays.
As well as these modes, one can exploit higher multiplicity decays, such as
$\btodzerokstar$, $\btodzerokpipi$~\cite{gronau} and $\bstodskpipi$.
Moreover, the decay $\bstodspipipi$ has been used to measure $\Delta m_s$~\cite{deltamslhcb}, 
and with a sufficiently large sample, provides a calibration for the 
flavor-mistag rate for the time-dependent analysis of $\bstodskpipi$.

\begin{figure}[ht]
\begin{center}
\includegraphics[width=65mm]{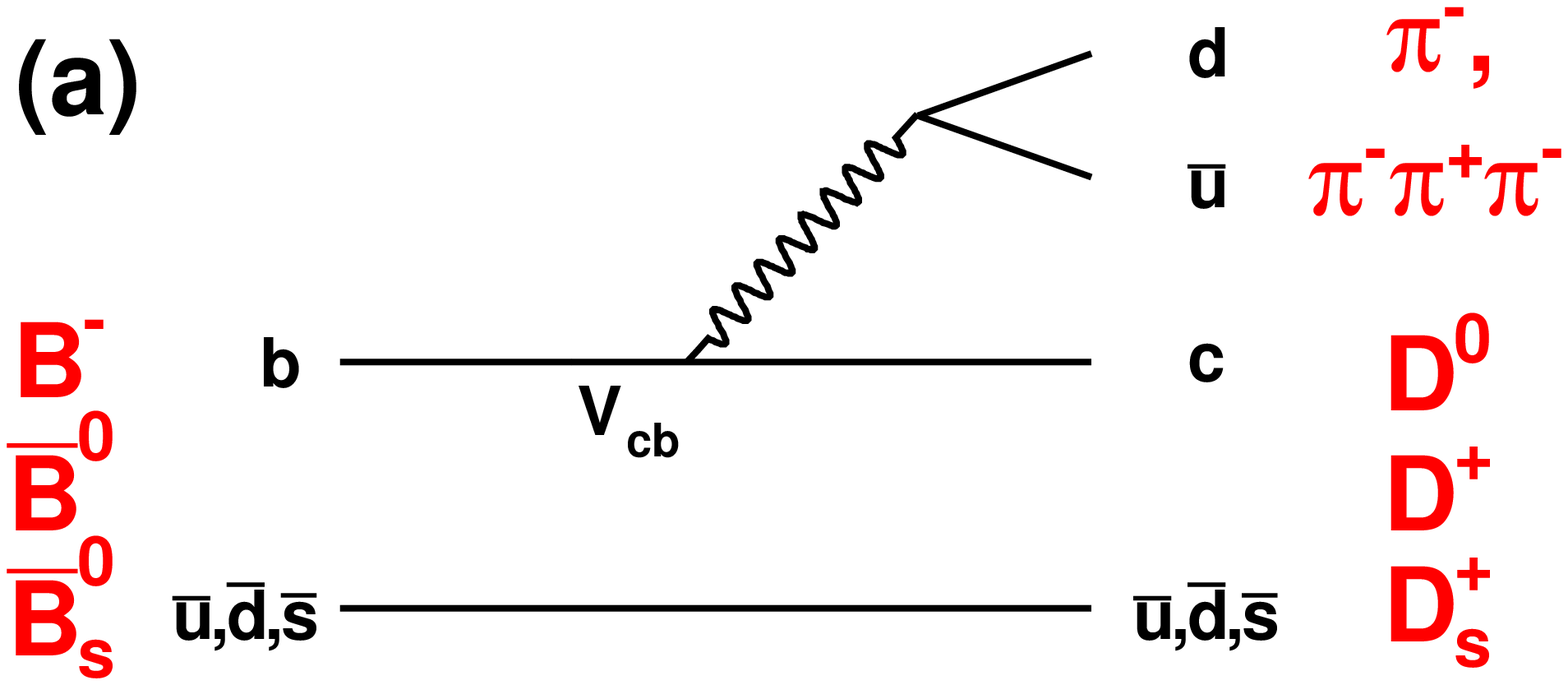}
\includegraphics[width=65mm]{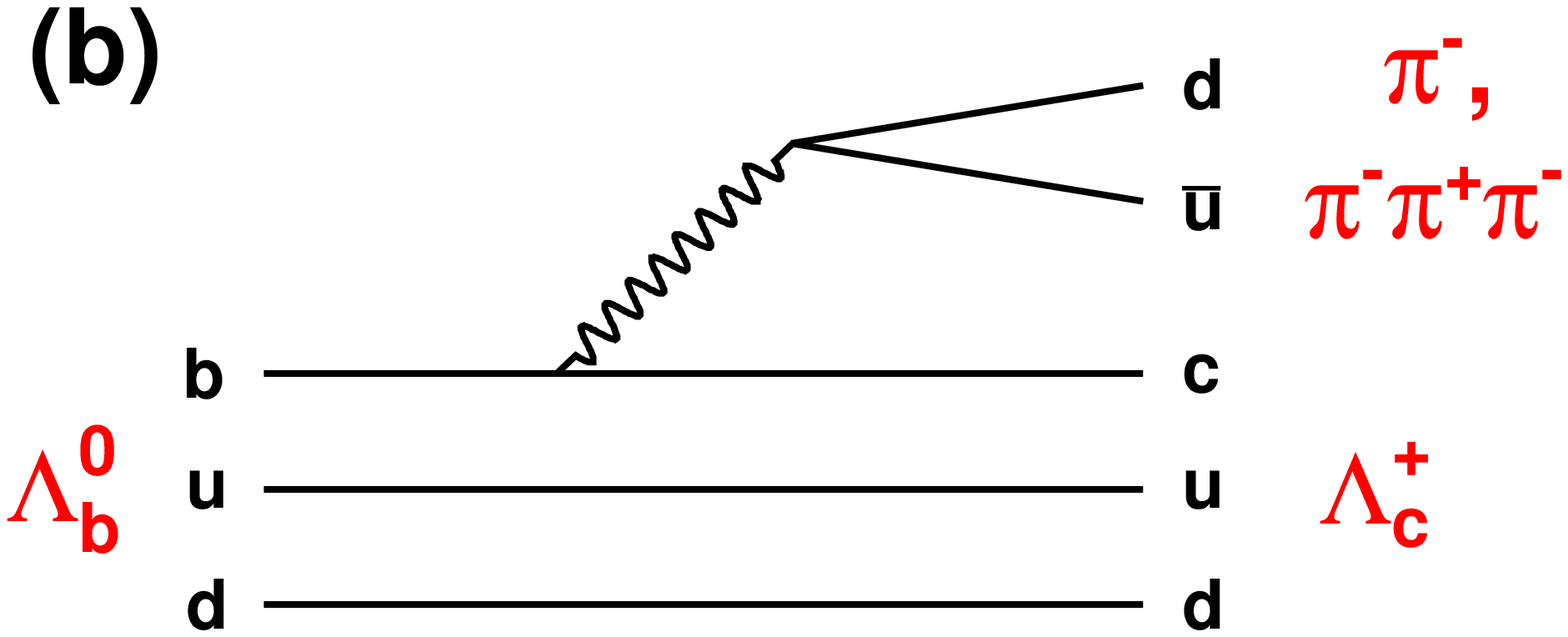}
\end{center}
\vspace{0.1in}
\begin{center}
\includegraphics[width=65mm]{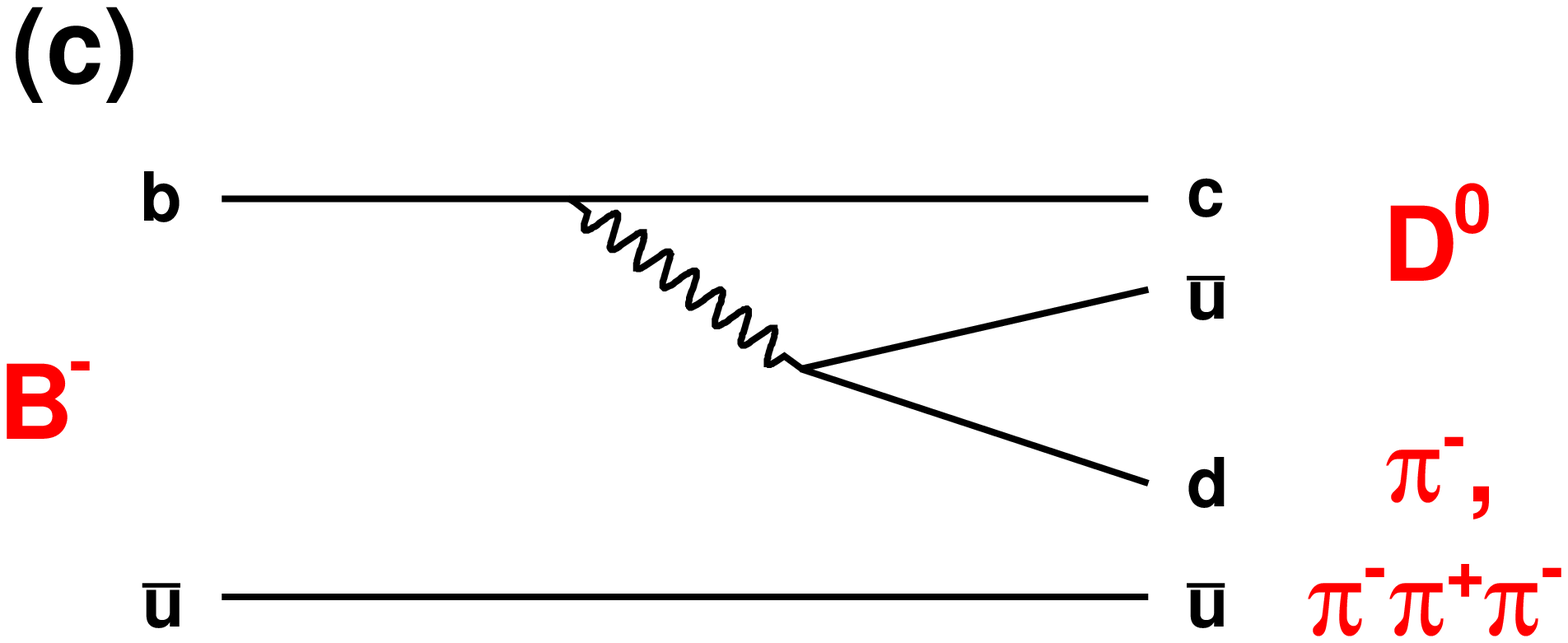}
\includegraphics[width=65mm]{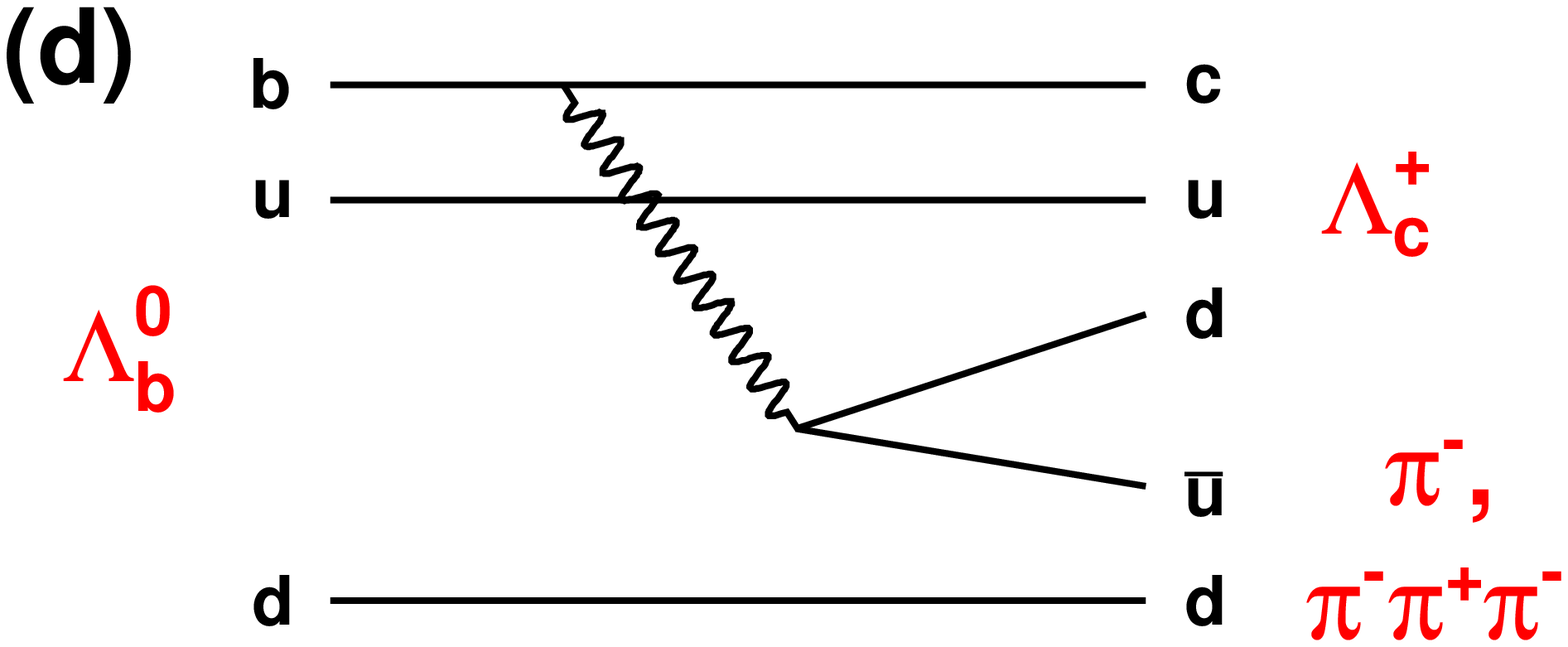}
\end{center}
\vspace{0.1in}
\begin{center}
\includegraphics[width=65mm]{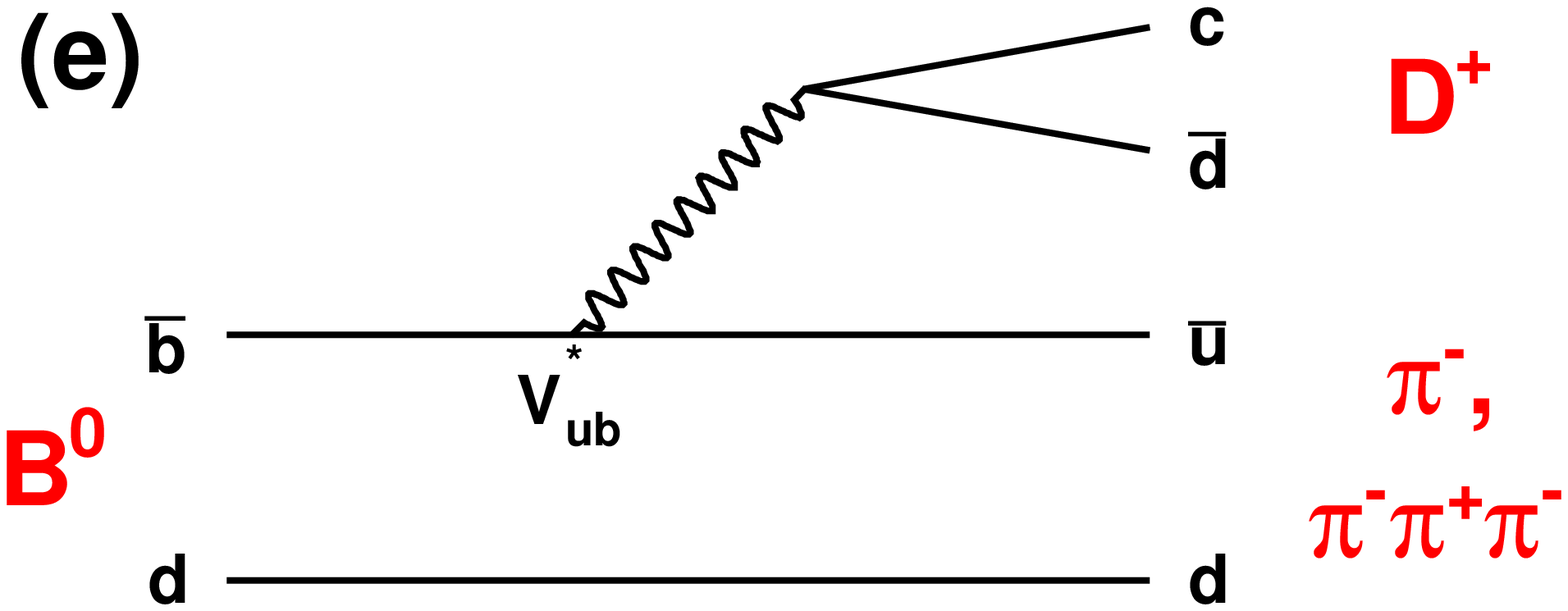}
\end{center}
\caption{Feynman diagrams for $\xbtoxcpi$ and $\xbtoxcpipipi$ decays. Figs. (a) and
(b) show external tree diagrams, (c) and (d) show color-suppressed tree diagrams
($B^-$ and $\Lambda_b^0$ only), and (e) shows the Cabibbo-suppressed external tree diagram,
only accessible to $B^0$ meson.}
\label{fig:feyn}
\end{figure}

The first step towards exploiting these multi-body decays is to observe them and 
quantify their branching fractions. The more interesting Cabibbo-suppressed decays
are $O(\lambda^3)$ in the Wolfenstein parameterization~\cite{wolf}, and therefore require
larger data samples. Here, we present measurements of the Cabibbo-favored
$\xbtoxcpipipi$ decays. The leading amplitudes contributing to these final states are shown in
Fig.~\ref{fig:feyn}. Additional contributions
from annihilation and $W$-exchange diagrams are suppressed and are not shown here.
Note that for the $B^-$ and $\Lambda_b^0$ decays, unlike the $\Bzb$ and $\Bsb$,
there is potential for interference between diagrams with similar magnitudes. In
Ref.~\cite{rosner}, it is argued that this interference can explain the larger
rate for $\btodzeropi$ compared to $\btodpi$. Thus, it is interesting to
see whether this is also true when the final state contains three pions.

In this paper, we report measurements of the $\xbtoxcpipipi$ branching fractions,
relative to $\xbtoxcpi$. We also
report on the partial branching fractions, $\xb\to\xc^*\pi^-,~\xc^*\to\xc\pi^+\pi^-$, 
where $\xb$ is either $\Bzb$, $B^-$, or $\Lb$, and $\xc^*$ refers to
$D_1(2420)^{+,0}$, $D_2^*(2460)^0$, $\Lambda_c(2595)^+$, or $\Lambda_c(2625)^+$. 
We also present results on the partial rates for $\Lb\to\Sc(2544)^{0,++}\pi^{\pm}\pi^{-}$. 
Charge conjugate final states are implied throughout.

\section{Detector and Trigger}
\label{sec:det}
The data used for this analysis were collected by the LHCb experiment during the 2010 data 
taking period and comprise about 35~$\ipb$ of integrated luminosity. LHCb has excellent capabilities to
trigger on and reconstruct bottom and charm hadrons.
The most important element of the detector for this analysis is a charged particle tracking
system that covers the forward angular region from about $15-350$~mrad and $15-250$~mrad in the horizontal and 
vertical directions, respectively. It
includes a 21 station, one-meter long array of silicon strip detectors (VELO) that come within 8 mm of the LHC beams,
a 4 Tm dipole magnetic field, followed by three multi-layer tracking stations (T-stations) downstream of the dipole magnet. 
Each T-station is composed of a four layer silicon strip detector (IT) in the high occupancy region near
the beam pipe, an eight layer straw tube drift chamber (OT) composed of 5~mm diameter straws outside this 
high occupancy region. Just upstream of the dipole magnet is a four-layer silicon strip detector (TT).  
Overall, the tracking system provides an impact parameter (IP)
resolution of $\sim16\mu$m + 30$\mu$m/$p_T$ (transverse momentum, $p_T$ in GeV/$c$), and a momentum resolution that ranges 
from $\sigma_p/p\sim0.4\%$ at 3 GeV/$c$ to $\sim0.6\%$ at 100 GeV/$c$. Two Ring Imaging Cherenkov Counters (RICH) 
provide a kaon identification
efficiency of $\sim$95\% for a pion fake rate of a few percent, integrated over the momentum 
range from 3$-$100 GeV/$c$. Downstream of the second RICH is a Preshower/Scintillating Pad Detector (PS/SPD), and 
electromagnetic (ECAL) and hadronic (HCAL) calorimeters.
Information from the ECAL/HCAL is used to form the hadronic triggers. Finally, a muon system consisting of five 
stations is used for triggering on and identifying muons.

To reduce the 40~MHz crossing rate to about 2~kHz for permanent storage,
LHCb uses a two-level trigger system. The first level of the trigger, L0, is hardware based and searches for either a
large transverse energy cluster ($E_T>3.6$~GeV) in the calorimeters, or a single high $p_T$ or di-muon pair 
in the muon stations. Events passing L0 are read out and sent to a large computing farm, where they are analyzed
using a software-based trigger. The first level of the software trigger, called HLT1, 
uses a simplified version of the offline software to apply tighter selections on charged particles based on their 
$p_T$ and minimal IP to any primary vertex (PV), defined as the location of the
reconstructed $pp$ collision(s). The HLT1 trigger relevant for this analysis 
searches for a single track with IP larger than 125~$\mu$m, $p_T>1.8$~GeV/$c$, $p>12.5$~GeV/$c$, along with other 
track quality requirements. Events that pass HLT1 are analyzed by a second software level, 
HLT2, where the event is searched for 2, 3, or 4-particle vertices that are consistent with
$b$-hadron decays. Tracks are required to have $p>5$~GeV/$c$, $p_T>0.5$~GeV/$c$ 
and IP $\chi^2$ larger than 16 to any PV, where the $\chi^2$ value is 
obtained assuming the IP is equal to zero. We also demand that at least one track has $p_T>1.5$~GeV/$c$, 
a scalar $p_T$ sum of the track in the vertex exceed 4~GeV/$c$, and that the corrected mass\footnote{The corrected mass is defined as 
$M_{\rm cor}=\sqrt{M^2+p_{\rm trans}^2}$, where $M$ is the invariant mass of the 2, 3 or 4-track
candidate (assuming the kaon mass for each particle), and $p_{\rm trans}$ is the momentum imbalance transverse to
the direction of flight, defined by the vector that joins the primary and secondary vertices.} 
is between 4 and 7 GeV/$c^2$. %
These HLT trigger selections each have an efficiency in the range of 80$-$90\% for events that pass typical 
offline selections for a large range of $B$ decays. 
A more detailed description of the LHCb detector can be found in Ref.~\cite{lhcb-det}.

Events with large occupancy are known to have intrinsically high backgrounds and to be slow to 
reconstruct. Therefore such events were suppressed by applying global event cuts (GECs) to hadronically
triggered decays. These GECs included a maximum of 3000 VELO clusters, 3000 IT hits, and 10,000 OT hits. In addition,
hadron triggers were required to have less than 900 or 450 hits in the SPD, depending on the specific
trigger setting. 

\section{Candidate Reconstruction and Selection}
\label{sec:recsel}

Charged particles likely to come from a $b$-hadron decay are first identified by requiring
that they have a minimum IP $\chi^2$ with respect to any PV of more than 9.
We also require a minimum transverse momentum, $p_T>300$~MeV/$c$, except for $\xbtoxcpipipi$ decays, 
where we allow (at most) one track to have $200<p_T<300$~MeV/$c$.
Hadrons are identified using RICH information by requiring the difference in log-likelihoods ($\Delta LL$) 
of the different mass hypotheses to satisfy
$\Delta LL(K-\pi)>-5$, $\Delta LL(p-\pi)>-5$ and $\Delta LL(K-\pi)<12$, for kaons, protons and pions, respectively.
These particle hypotheses are not mutually exclusive, however the same track cannot 
enter more than once in the same decay chain.

Charm particle candidates are reconstructed in the decay modes $D^0\to K^-\pi^+$, 
$D^+\to K^-\pi^+\pi^+$, $D_s^+\to K^+K^-\pi^+$ and $\Lambda_c^+\to pK^-\pi^+$. The
candidate is associated to one of the PVs in the event based on the smallest IP $\chi^2$
between the charm particle's reconstructed trajectory and
all PVs in the event. A number of selection criteria are imposed to reduce backgrounds
from both prompt charm with random tracks as well as purely combinatorial background.
To reduce the latter, we demand that each candidate is well separated from the 
associated PV by requiring that its flight distance (FD) projected onto the $z$-axis is larger than 2~mm, 
the FD~$\chi^2>49$\footnote{This is the $\chi^2$ with respect to the FD=0 hypothesis.}, 
and that the distance in the transverse direction ($\Delta R$) is larger than 100~$\mu$m. 
Background from random track combinations is also suppressed by requiring the vertex fit 
$\chi^2$/ndf$<8$, and $p_T>1.25$~ GeV/$c$ (1.5 GeV/$c$ for $D^+_{(s)}$ in $\bstodspi$.)
To reduce the contribution from prompt charm, we require that the charm particle has 
a minimal IP larger than 80~$\mu$m and IP $\chi^2>12.25$ with respect to its associated PV.
For $D_s^+\to K^+K^-\pi^+$, we employ tighter particle identification requirements on the
kaons, namely $\Delta LL(K-\pi)>0$, if the $K^+K^-$ invariant mass is outside a window of $\pm$20~MeV/$c^2$ 
of the $\phi$ mass~\cite{pdg}. 
Lastly, we require the reconstructed charm particles masses to be within 25~MeV/$c^2$ of their known values. 

The bachelor pion for $\xbtoxcpi$ is required to have $p_T>0.5$~GeV/$c$,
$p>5.0$~GeV/$c$ and IP $\chi^2>16$. For the 3$\pi$ vertex associated with the
$\xbtoxcpipipi$ decays, we apply a selection identical to that for the charm particle candidates, except we only
require the $p_T$ of the $3\pi$ system to be larger than 1~GeV/$c$ and that the invariant mass is in the range from 
$0.8~{\rm GeV}/c^2<M(\pi\pi\pi)<3.0$~GeV/$c^2$.

Beauty hadrons are formed by combining a charm particle with either a single pion 
candidate (for $\xbtoxcpi$) or a $3\pi$ candidate (for $\xbtoxcpipipi$.) The $b$-hadron 
is required to have a transverse momentum of at least 1~GeV/$c$. As with the
charm hadron, we require it is well-separated from its associated PV, with
FD larger than 2~mm, FD $\chi^2>49$ and $\Delta R>100$~$\mu$m. We also make a series of requirements that ensure
that the $b$-hadron candidate is consistent with a particle produced in a proton-proton interaction. We
require the candidate to have IP$<$90~$\mu$m, IP $\chi^2 <16$, and 
that the angle $\theta$ between the $b$-hadron momentum and the vector formed by joining the
associated PV and the decay vertex satisfies $\cos\theta>0.99996$. 
To ensure a good quality vertex fit, we require a vertex fit $\chi^2/$ndf$<6$ (8 for $\xbtoxcpi$.)

To limit the timing to process high occupancy events, we place requirements on the number of 
tracks\footnote{Here, tracks refer to charged particles that have segments in both the
VELO and the T-stations.} in an event. For $\btodpi$ and $\bstodspi$, the maximum number of 
tracks is 180, and for $\LbtoLcpi$ and $\btodzeropi$ it is 120. These selections are 99\% and 95\% efficient,
respectively, after the GECs. The $\xbtoxcpipipi$ selection requires fewer than 300 tracks, and thus is
essentially 100\% efficient after the GECs.

Events are required to pass the triggers described above.
This alone does not imply that the signal $b$-hadron decay was directly responsible for the trigger.
We therefore also require that one or more of the signal $b$-hadron daughters is responsible for triggering the event.
We thus explicitly select events that {\bf T}riggered {\bf O}n the {\bf S}ignal 
decay ({\bf TOS}) at L0, HLT1 and HLT2. For the measurements of excited charm states, where our yields
are statistically limited, we also make use of 
L0-triggers that {\bf T}riggered {\bf I}ndependently of the {\bf S}ignal decay ({\bf TIS}).
In this case, the L0 trigger is traced to one or more particles other than those in the signal decay.

Lastly, we note that in $\xbtoxcpipipi$ candidate events, between 4\% and 10\% have multiple candidates
(mostly two) in the same event. In such cases we choose the candidate with the largest transverse momentum. This 
criterion is estimated to be $(75\pm20)$\% efficient for choosing the correct candidate. For $\xbtoxcpi$ multiple
candidates occur in less than 1\% of events, from which we again choose the one with the largest $p_T$.

\subsection{Selection Efficiencies}

Selection and trigger efficiencies are estimated using Monte Carlo (MC) simulations. The MC samples
are generated with an average number of interactions per crossing equal to 2.5, which is similar to the running
conditions for the majority of the 2010 data. The $b$-hadrons are produced
using {\sc pythia}~\cite{pythia} and decayed using {\sc evtgen}~\cite{evtgen}. 
The $\xbtoxcpipipi$ decays are produced using a cocktail for the $\pi\pi\pi$ system
that is $\sim$2/3 $a_1(1260)^-\to\rho^0\pi^-$ and about 1/3 non-resonant $\rho^0\pi^-$. 
Smaller contributions from $D_1^{0}(2420)\pi$ and
$D_2^{*0}(2460)\pi$ are each included at the 5\% level to $\btodzeropipipi$ and 2\% each for $\btodpipipi$. 
For $\LbtoLcpipipi$, we include
contributions from $\Lambda_c(2595)^+$ and $\Lambda_c(2625)^+$, which contribute 9\% and 7\% to the 
MC sample. The detector is simulated with {\sc geant4}~\cite{geant}, 
and the event samples are subsequently analyzed in the same way as data. 

We compute the total kinematic efficiency, $\eff_{\rm kin}$ from the MC simulation as the fraction of
all events that pass all reconstruction and
selection requirements. These selected events are then passed through a software emulation of the L0 trigger, 
and the HLT software used to select the data, from which we compute the trigger efficiency ($\epsilon_{\rm trig}$).
The efficiencies for the decay modes under study are shown in
Table~\ref{tab:kineff}.  Only the relative efficiencies are used to obtain the results in this paper.

\begin{table*}[ht]
\begin{center}
\caption{Summary of efficiencies for decay channels under study. Here, $\eff_{\rm kin}$ is the total 
kinematic selection efficiency, $\eff_{\rm trig}$ is the trigger efficiency, and $\eff_{\rm tot}$ is
their product. The uncertainties shown are statistical only.}
\begin{tabular}{lccc}
\hline\hline
Decay & $\eff_{\rm kin}$ & $\eff_{\rm trig}$ & $\eff_{\rm tot}$ \\
      &        (\%)      &        (\%)      &        (\%)     \\
\hline
\raisebox{-0.5ex}{$\btodpipipi$} &  \raisebox{-0.5ex}{$0.153\pm0.003$} & \raisebox{-0.5ex}{$22.6\pm0.5$} &  \raisebox{-0.5ex}{$0.0347\pm0.0011$} \\ 
$\btodzeropipipi$ & $0.275\pm0.007$ &  $27.4\pm0.6$  &  $0.0753\pm0.0019$\\
$\bstodspipipi$ & $0.137\pm0.003$ & $24.9\pm0.7$ &  $0.0342\pm0.0012$\\
$\LbtoLcpipipi$ & $0.110\pm0.005$ & $24.0\pm0.7$ &  $0.0264\pm0.0008$ \\ [1ex]
$\btodpi$ &  $0.882\pm0.014$ &  $20.8\pm0.3$  &  $0.184\pm0.004$  \\
$\btodzeropi$ & $1.54\pm0.02$ & $27.4\pm0.3$ &  $0.421\pm0.007$ \\
$\bstodspi$ & $0.868\pm0.010$  &  $23.1\pm0.2$  &  $0.201\pm0.003$\\
$\LbtoLcpi$ & $0.732\pm0.015$ &  $24.7\pm0.4$  &  $0.181\pm0.004$ \\
\hline\hline
\end{tabular}
\label{tab:kineff}
\end{center}
\end{table*}

\section{Reconstructed Signals in Data}
\label{sec:signals}

\begin{figure}[ht]
\centering
\includegraphics[width=75mm]{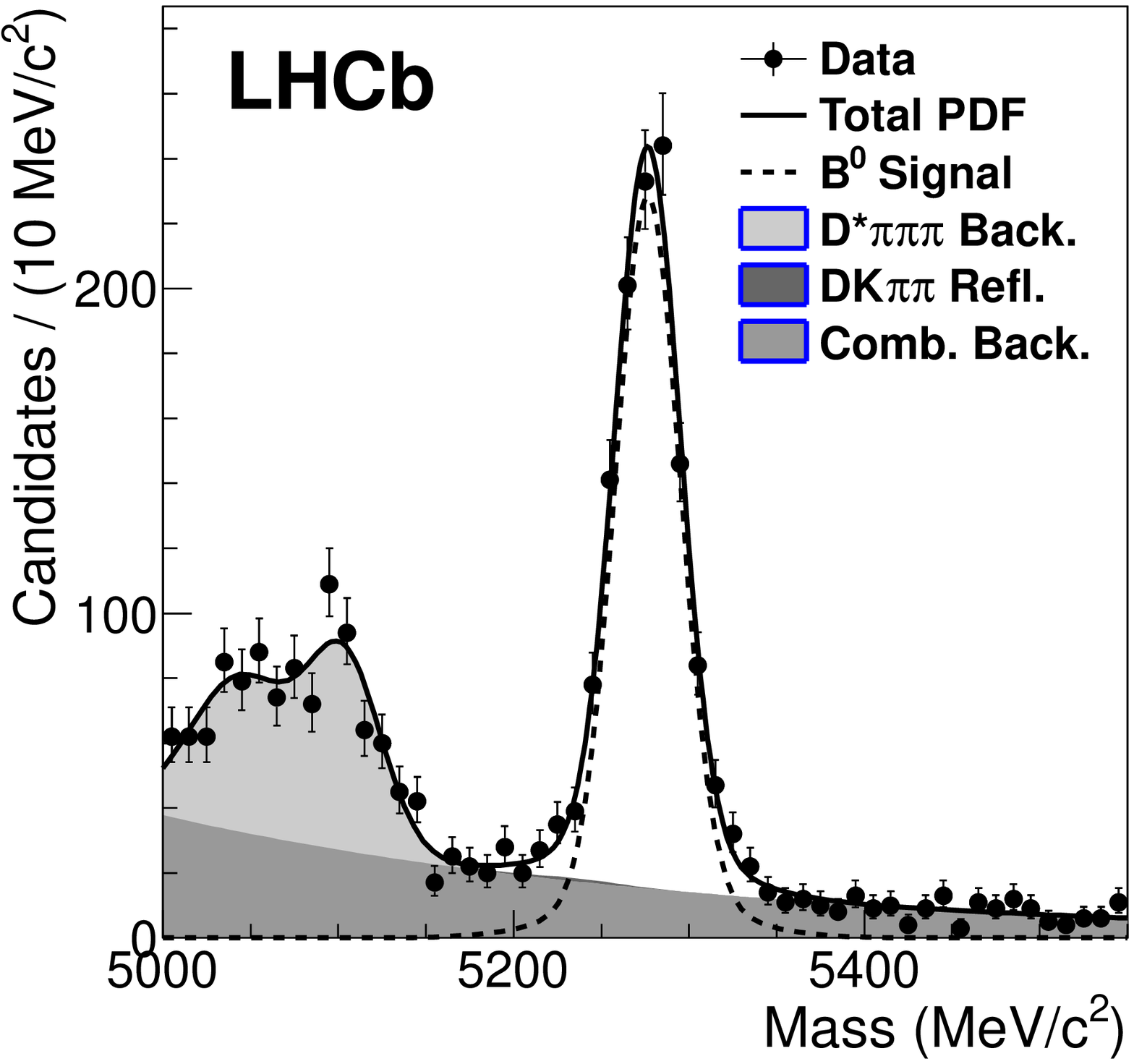}
\includegraphics[width=75mm]{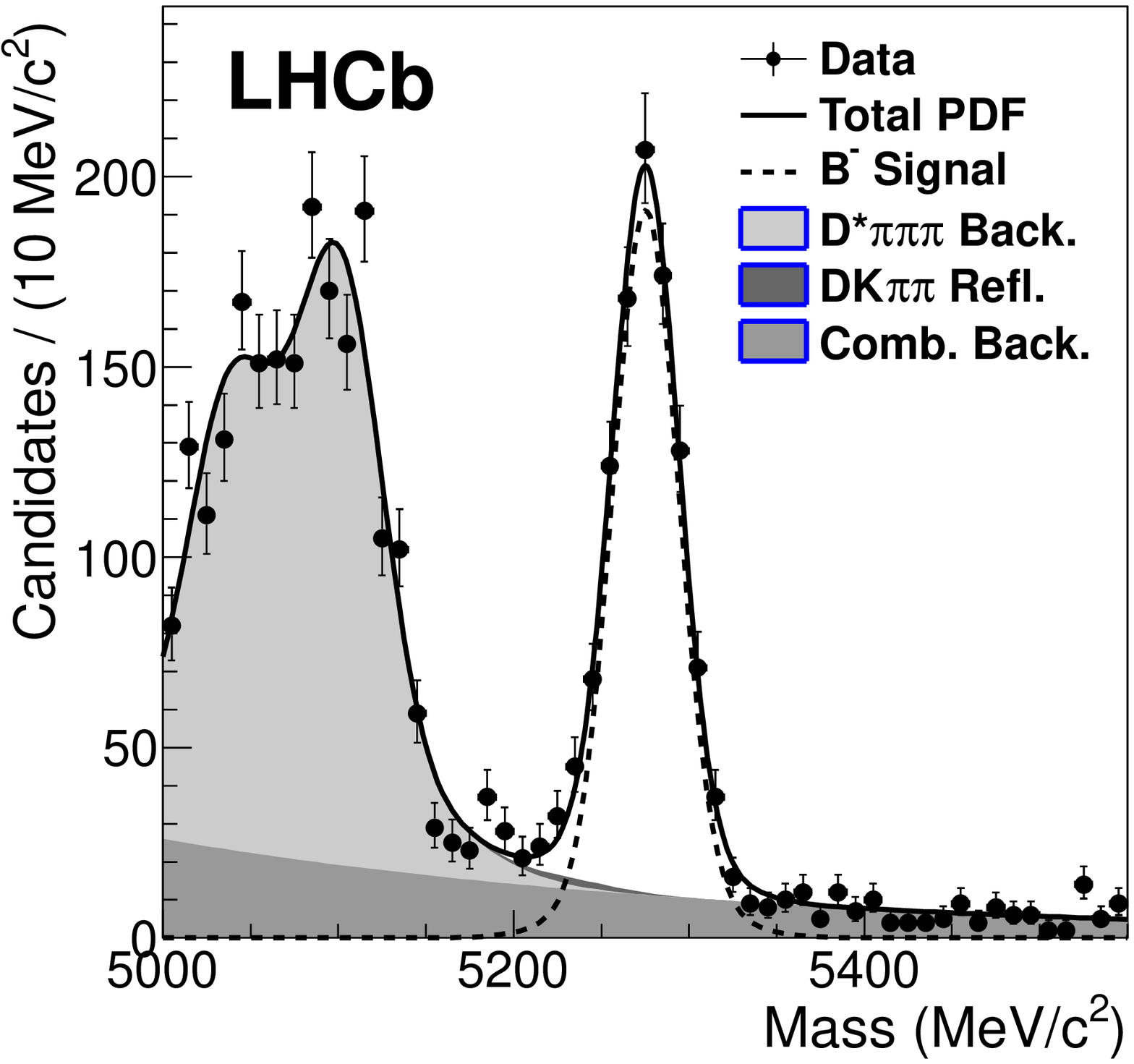}
\includegraphics[width=75mm]{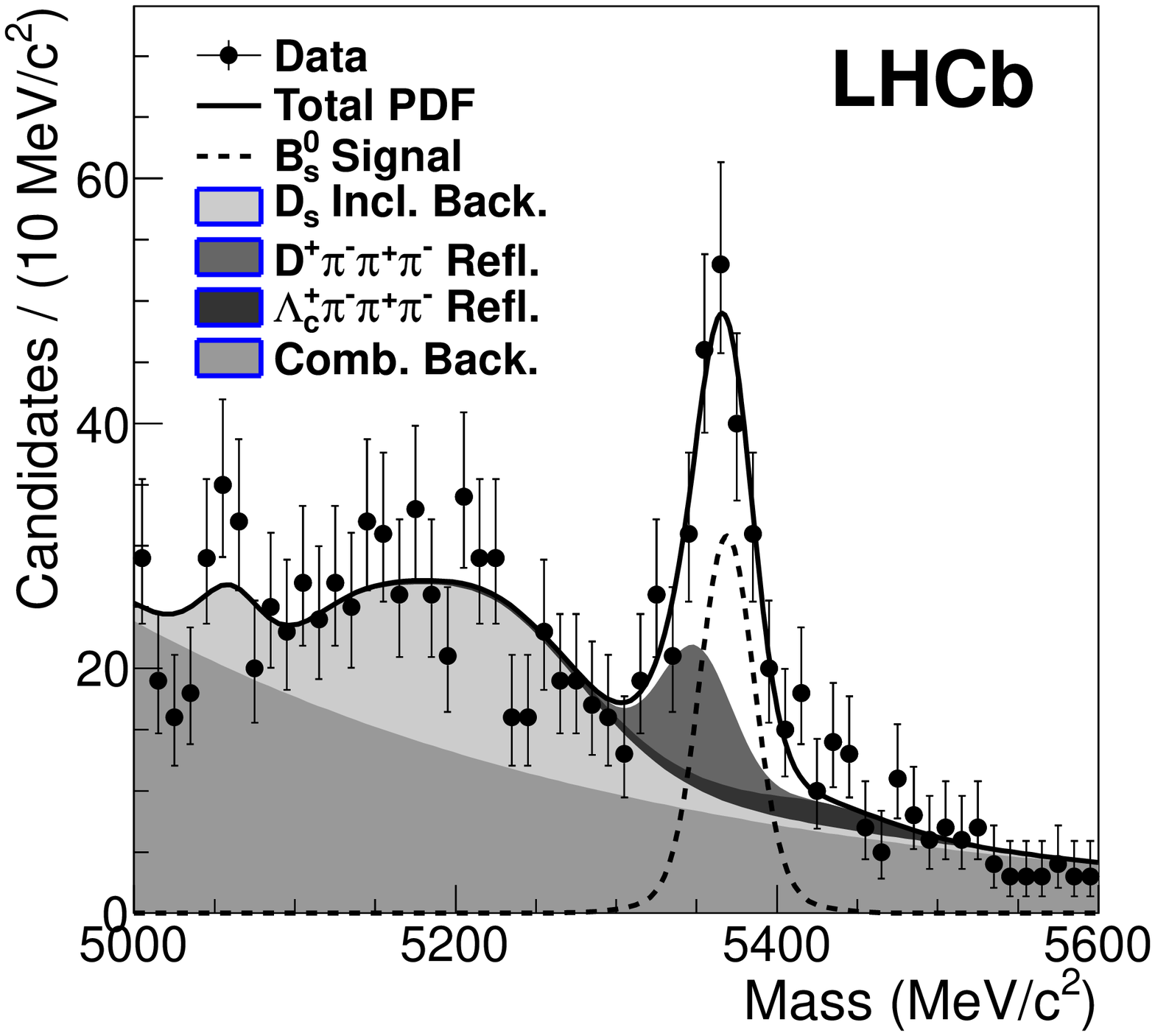}
\includegraphics[width=75mm]{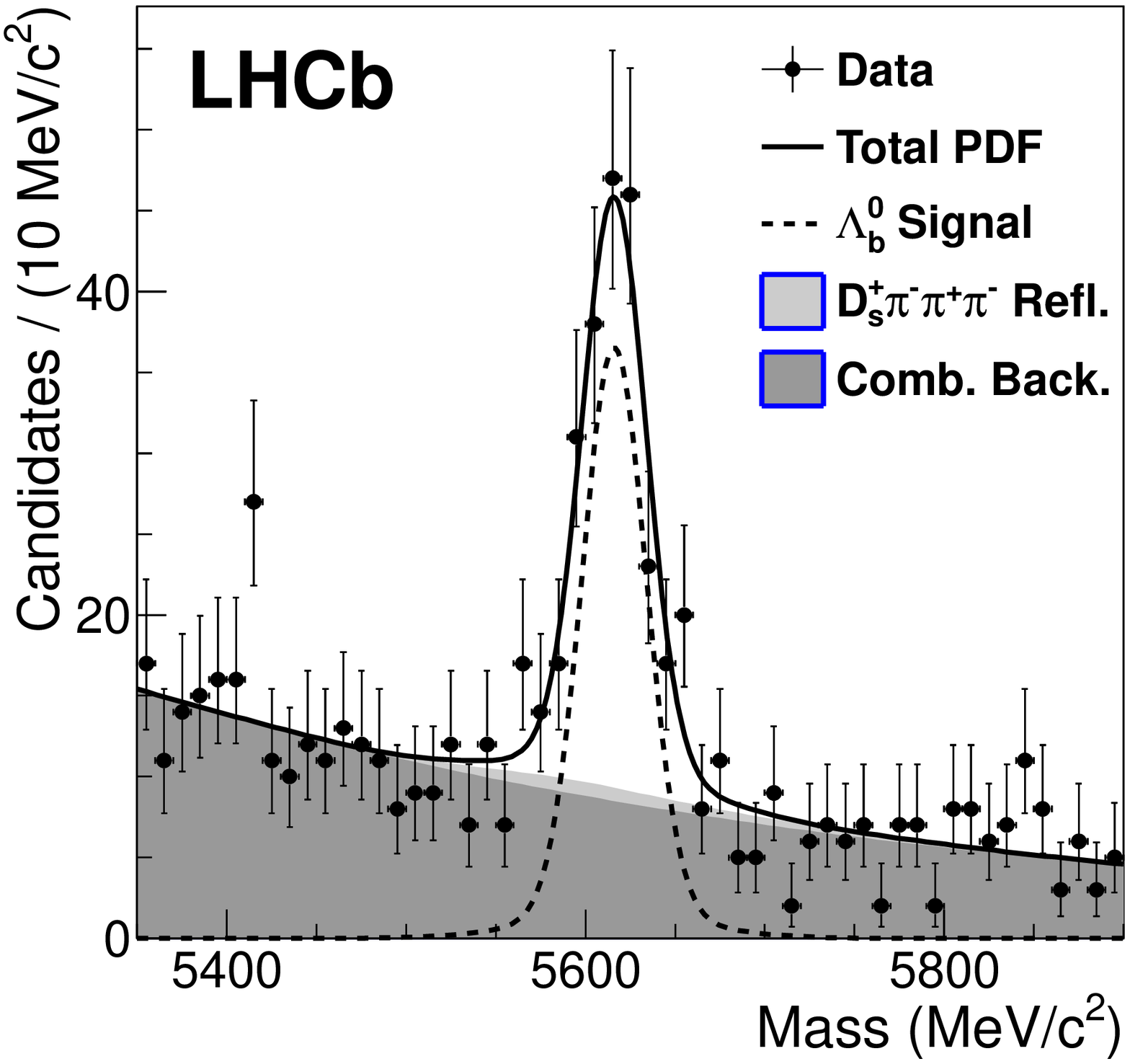}
\caption{Invariant mass distributions for $\btodpipipi$ (top left), $\btodzeropipipi$ (top right),
$\bstodspipipi$ (bottom left), and $\LbtoLcpipipi$ (bottom right). Fits showing the signal and
background components are indicated, and are described in the text.}
\label{fig:xbtoxc3pi}
\end{figure}

\begin{figure}[ht]
\centering
\includegraphics[width=75mm]{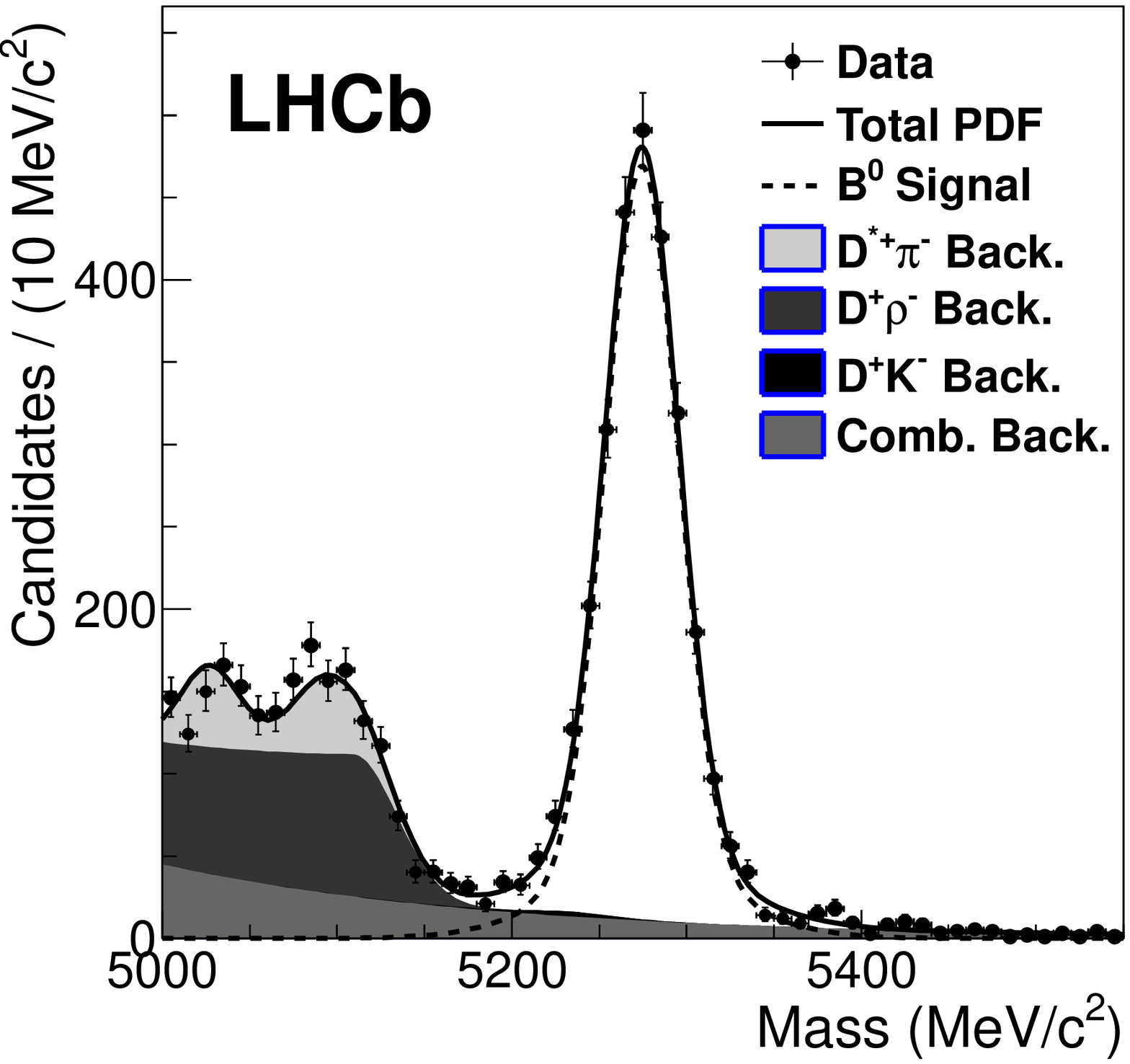}
\includegraphics[width=75mm]{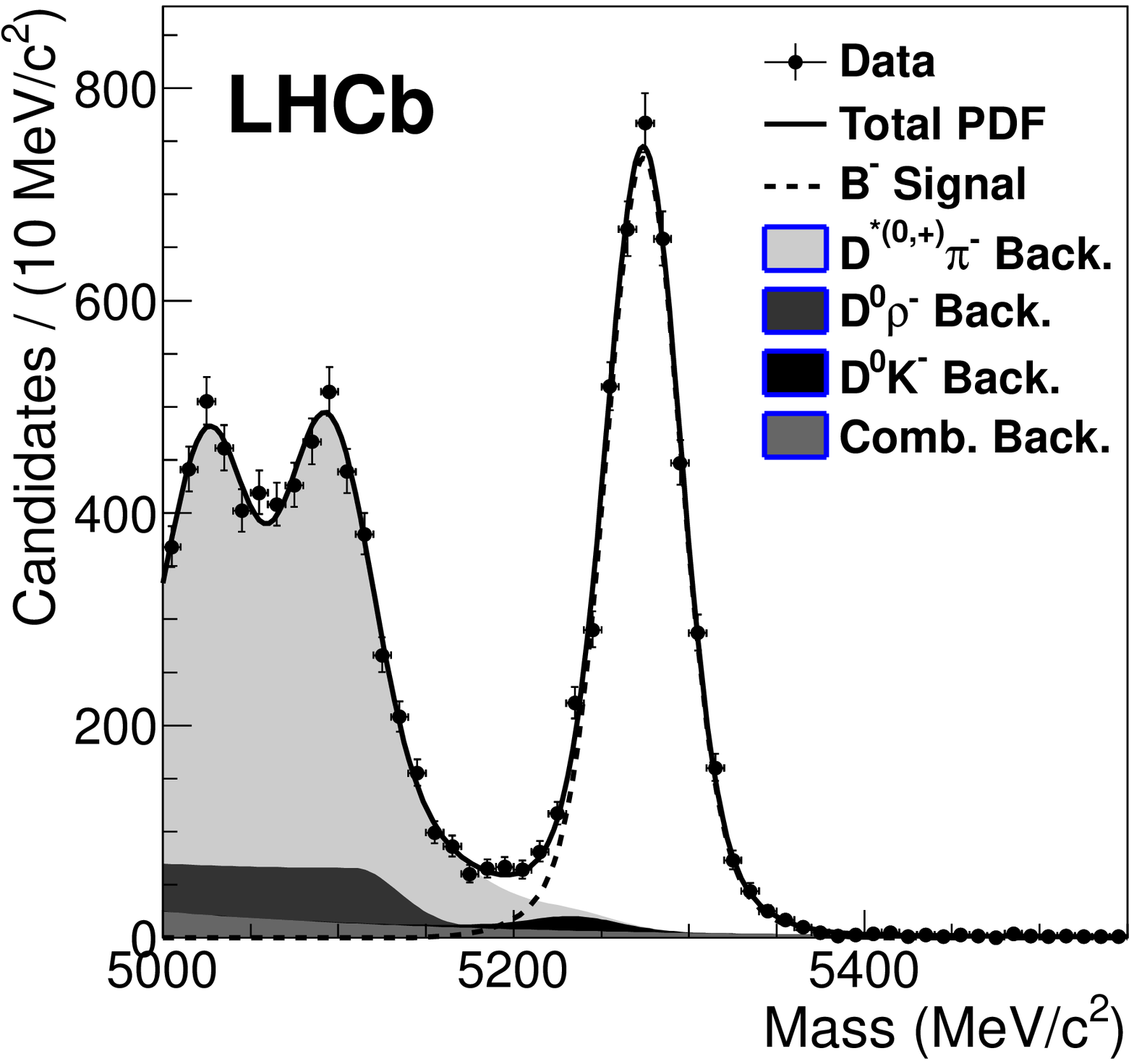}
\includegraphics[width=75mm]{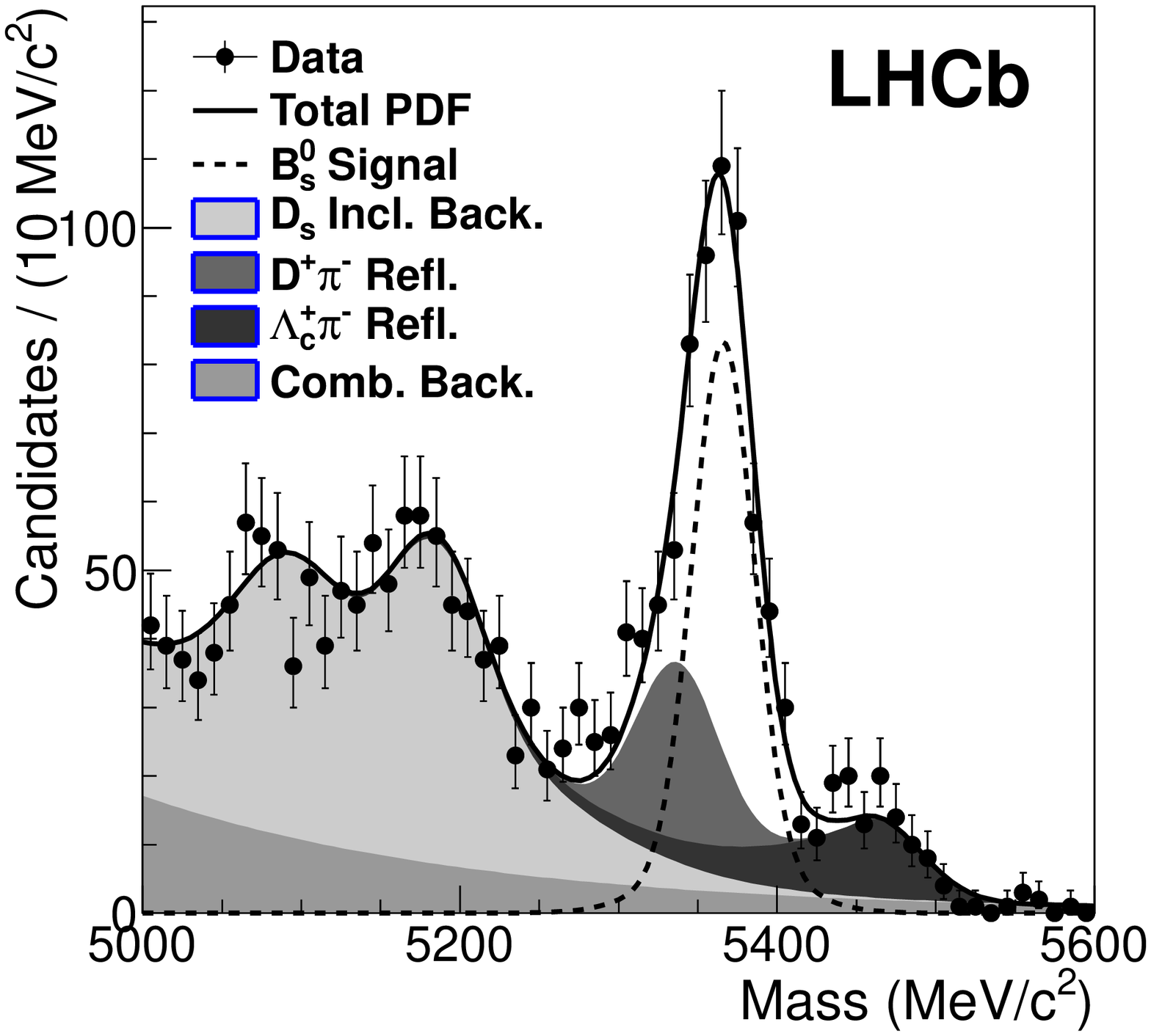}
\includegraphics[width=75mm]{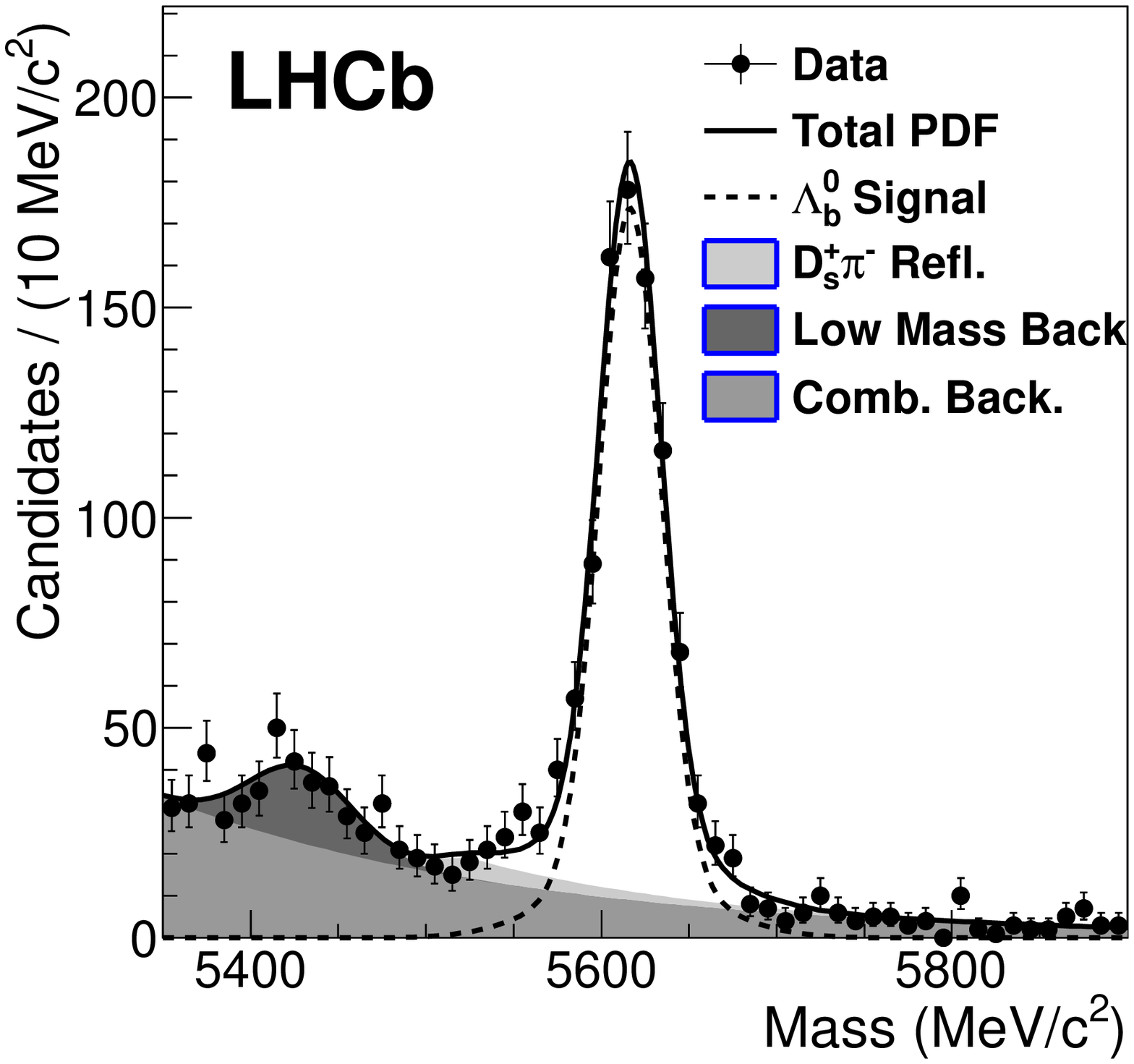}
\caption{Invariant mass distributions for $\btodpi$ (top left), $\btodzeropi$ (top right),
$\bstodspi$ (bottom left), and $\LbtoLcpi$ (bottom right). Fits showing the signal and
background components are indicated, and are described in the text.}
\label{fig:xbtoxcpi}
\end{figure}

The reconstructed invariant mass distributions 
are shown in Figs.~\ref{fig:xbtoxc3pi} and~\ref{fig:xbtoxcpi} for the signal and normalization
modes, respectively. Unbinned likelihood fits are performed to extract the signal yields, where the
likelihood functions are given by the sums of signal and several background components. The signal and background
components are shown in the figures. The signal contributions are each described by the sum of two Gaussian shapes 
with equal means. The relative width and fraction of the wider Gaussian shape with respect to the narrower one
are constrained to the values found from MC simulation based on agreement with data in the large yield signal modes.
This constraint is included with a 10-12\% uncertainty (mode-dependent), which is the level of agreement found between data and
MC simulation. The absolute width of the narrower Gaussian is a free parameter in the fit, since the data shows a slightly
worse ($\sim$10\%) resolution that MC simulation.

For  $\bstodspi$ and $\bstodspipipi$ decays, there are peaking backgrounds from $\btodpi$ and $\btodpipipi$
just below the $B_s^0$ mass. We therefore fix their core Gaussian widths as well, based on the resolutions 
found in data for the kinematically similar $\btodpi$ and $\btodpipipi$ decays, scaled by
0.93, which is the ratio of expected widths obtained from MC simulation. 

A number of backgrounds contribute to these decays. Below the $b$-hadron masses there are generally 
peaking background structures due to partially reconstructed $B$ decays. These decays include
$B_{(s)}\to D_{(s)}^*\pi(\pi\pi)$, with a missed photon, $\pi^0$ or $\pi^+$, as well as $B_{(s)}\to D_{(s)}\rho^-$, 
where the $\pi^0$ is not included in the decay hypothesis.
For the $\btodpi$ and $\btodzeropi$ decays, the shapes of these backgrounds are taken from dedicated signal MC 
samples. The double-peaked background shape from partially-reconstructed $D^*\pi$ decays is obtained
by fitting the background MC sample to the sum of two Gaussian shapes with different means. The difference
in their means is then fixed, while their average is a free parameter in subsequent
fits to the data. For $\btodpipipi$ and $\btodzeropipipi$, the shape of the partially-reconstructed 
$D^*\pi\pi\pi$  background is not as easily derived since the helicity amplitudes are not known. 
This low mass background is also parametrized using a two-Gaussian model, but we let the parameters float
in the fit to the data. For $\bstodspi$ and $\bstodspipipi$,
we obtain the background shape from a large $\Bsb\to D_s^+ X$ inclusive MC sample. Less is 
known about the $\Lb$ hadronic decays that would contribute background to the $\LbtoLcpi$ and $\LbtoLcpipipi$
invariant mass spectra. For $\LbtoLcpipipi$, we see no
clear structure due to partially-reconstructed backgrounds. For $\LbtoLcpi$, there
does appear to be structure at about 5430~MeV/$c^2$, which may be due to $\Lambda_c^+\rho^-$.
The enhancement is described by a single Gaussian above the combinatoric background,
which, given the limited number of events, provides a good description of this background.

There are also so-called reflection backgrounds, where fully reconstructed signal decays from one $b$-hadron
decay mode produce peaking structures in the invariant mass spectra of other decay modes when
one of the daughter particles is misidentified. For  $B\to D\pi^-(\pi^+\pi^-)$, there 
are reflections from $B\to DK^-(\pi^+\pi^-)$  Cabibbo-suppressed decays, where the
kaon is misidentified as a pion. Due to the Cabibbo 
suppression and the excellent RICH performance, their contributions are limited to the 1\% level. 
The shape of this misidentification background is taken from MC simulation and is constrained to be
$(1\pm1)\%$ of the signal yield. 

For the $\bstodspi$ and $\bstodspipipi$ decays, there are reflection backgrounds from 
$\btodpi$ and $\btodpipipi$ modes, when either of the $\pi^+$ from the $D^+$ decay is misidentified as a $K^+$.
This cross-feed background is evaluated in two
ways. First, we take our $\btodpi$ ($\btodpipipi$) data, which have very loose particle identification (PID)
requirements on the pions, and apply the kaon PID selection to them. If either of the two pions pass, 
and the recomputed ($KK\pi$) mass is within the $D_s^+$ mass window, the candidate is counted as a reflection background.
Using this technique, we find $(5.3\pm0.4)\%$ [$(6.3\pm0.6)\%$] of $\btodpi$ 
[$\btodpipipi$] signal decays reflect into the $\bstodspi$ [$\bstodspipipi$] signal region.
In the second method, we apply a $\pi$-faking-$K$ misidentification matrix (in bins of $p$ and $p_T$),
obtained from a $D^{*+}$ data calibration sample to the $\btodpi$ (or $\btodpipipi$) signal MC sample, 
followed by the $D_s^+$ mass window requirement (after replacing the pion mass with the kaon mass.)
The results of this second procedure are $(4.4\pm0.3)\%$ for $\btodpi$ and $(5.2\pm0.4)\%$ for $\btodpipipi$,
both of which are consistent with the first method. We therefore constrain the peaking background from
$\btodpi$ ($\btodpipipi$) into $\bstodspi$ ($\bstodspipipi$) to be $(4.0\pm1.5)\%$ ($(5.0\pm2.0)\%$),
where the Gaussian constraint is conservatively assigned a 40\% relative uncertainty.
The shape of this peaking background is obtained from MC simulation and is well-described by a single
Gaussian of mean 5350~MeV/$c^2$ and width 30~MeV/$c^2$. This shape is in good agreement with what is 
observed in data.

The second reflection background to $\bstodspi$ ($\bstodspipipi$) is $\LbtoLcpi$ ($\LbtoLcpipipi$), 
where the proton from the $\Lambda_c$ decay is misidentified as a kaon. This is similar to the
$\Bzb$ reflection, except here the $\Lambda_b^0$ yield is significantly smaller, obviating
the need for making an explicit $\Delta LL(K-p)$ requirement to reject protons.
The $\Lambda_b^0$ reflection background is evaluated using the first technique as described above
leading to reflection rates of $(15\pm3)\%$ for $\LbtoLcpi$ into $\bstodspi$ and $(20\pm4)\%$ for $\LbtoLcpipipi$
into $\bstodspipipi$. We conservatively assign a 20\% uncertainty on this rate based on the agreement between
data and MC simulation. The asymmetric shape of this background is described by the simulation, which is consistent
with the shape observed in data.
The combinatorial background is modeled with an exponential distribution. The fits are superimposed on the data in
Figs.~\ref{fig:xbtoxc3pi} and~\ref{fig:xbtoxcpi}, and the fitted yields are summarized in Table~\ref{tab:brinputs}.

\noindent 

The ratios of branching ratios are given by:

\begin{align}
  {\br(\xbtoxcpipipi)\over \br(\xbtoxcpi) } = {Y^{\rm sig}/\eff_{\rm tot}^{\rm sig}
    \over Y^{\rm norm}/\eff_{\rm tot}^{\rm norm} } 
  \nonumber
\end{align}

\noindent where the $Y$ factors are the observed yields in the signal 
and normalization modes, and $\eff_{\rm tot}$ are the total selection efficiencies. 

\begin{table*}[ht]
\begin{center}
\caption{Summary of yields for the branching fraction computation. Uncertainties are statistical only.}
\begin{tabular}{lclc}
\hline\hline
Decay             &     Yield     &   Decay          &     Yield  \\
\hline 
\raisebox{-0.5ex}{$\btodpipipi$}     &   \raisebox{-0.5ex}{$1150\pm43$} & \raisebox{-0.5ex}{$\btodpi$ }    & \raisebox{-0.5ex}{$2745\pm66$} \\
$\btodzeropipipi$ & $950\pm41$    & $\btodzeropi$ & $4244\pm90$ \\
$\bstodspipipi$   & $138\pm23$    & $\bstodspi$   & $434\pm32$ \\
$\LbtoLcpipipi$   & $174\pm18$    & $\LbtoLcpi$   & $853\pm36$ \\
\hline\hline
\end{tabular}
\label{tab:brinputs}
\end{center}
\end{table*}

\section{Systematic Uncertainties}
\label{sec:syst}

Several sources contribute uncertainty to the measured ratios of branching fractions. Because
we are measuring ratios of branching fractions, most, but not all of the potential systematics cancel.
Here, we discuss only the non-cancelling uncertainties. With regard to the reconstruction
of the $\xbtoxcpipipi$ and $\xbtoxcpi$ decays, the former has two additional pions which need to
pass our selections, and the $3\pi$ system needs to pass the various vertex-related selection criteria.
The track reconstruction efficiency and uncertainty are evaluated by measuring the ratio of 
fully reconstructed $J/\psi$'s to all $J/\psi$'s obtained from an inclusive single muon trigger,
where only one of the muons is required to be reconstructed.
After reweighting the efficiencies to match the kinematics of the signal tracks, the uncertainty is
found to be 2\% per track, which leads to a 4\% uncertainty in the branching
fraction ratios. The IP resolution in data is about 20\% worse than in the simulation, leading
to (i) a larger efficiency for tracks to pass the IP-related cuts (as well as larger background),
and (ii) a lower efficiency to pass the vertex $\chi^2$ selections, for data relative to the
value predicted by simulation. The first of these is studied by reducing the IP $\chi^2$ requirement
in simulation by 20\%, and the second by smearing the vertex $\chi^2$ distribution in
simulation until it agrees with data. The combined correction is found to be $1.02\pm0.03$.

Another potential source of systematic uncertainty is related to the production and decay model for
producing the $\xc\pi\pi\pi$ final state. We have considered that the $p_T$ spectrum of the pions
in the 3$\pi$ system may be different between simulation and data. To estimate the uncertainty, we reweight the
MC simulation to replicate the momentum spectrum of the lowest momentum pion (amongst the pions 
in the $3\pi$ vertex.) We find that the total efficiency
using the reweighted spectra agrees with the unweighted spectra to within 3\%. We have
also investigated the effect of differences in the $p_T$ spectra of the charm particle,
and find at most a 1\% difference. Our candidate selection is limited to the mass region 
$M(\pi\pi\pi)<3$~GeV/$c^2$. Given that the 
phase space population approaches zero as $M(\pi\pi\pi)\to 3.5$~GeV/$c^2$ ({\it i.e.,} $M_B-M_D$)
and that the simulation reasonably reproduces the $\pi^-\pi^+\pi^-$ mass spectrum, we
use the simulation to assess the fraction of the $\pi\pi\pi$ mass spectrum beyond 3~GeV/$c^2$.
We find the fraction of events above 3~GeV/$c^2$ is $(3.5-4.5)$\% for the decay modes under study.
We apply a correction of $1.04\pm0.02$, where we have assigned half the correction as 
an estimate of the uncertainty. In total, the correction for production and decay models is $1.04\pm0.04$.

As discussed in Sec.~\ref{sec:recsel}, we choose only one candidate per event. The efficiency of
this selection is estimated by comparing the signal yield in multiple-candidate events before and
after applying the best candidate selection. The selection is estimated to be $(75\pm20)\%$ efficient.
In the $\xbtoxcpipipi$ the multiple candidate rate varies from 4\% to 10\%,
so we have corrections that vary from 1.01 to 1.03. For $\xbtoxcpi$, this effect is negligible.
The corrections for each mode are given in Table~\ref{tab:syst}.

For the trigger efficiency, we rely on signal MC simulations to emulate the online trigger.
The stability of the relative trigger efficiency was checked by reweighting the $b$-hadron
$p_T$ spectra for both the signal and normalization modes, and re-evaluating the trigger efficiency ratios. 
We find maximum differences of 2\% for L0, 1\% for HLT1 and 1\% for HLT2,
(2.4\% total) which we assign as a systematic uncertainty.

Fitting systematics are evaluated by varying the background shapes and assumptions about the signal
parameterization for both the $\xbtoxcpipipi$ 
and $\xbtoxcpi$ modes and re-measuring the yield ratios. For the combinatorial background,
using first and second order polynomials leads to a 3\% uncertainty on the relative yield.
Reflection background uncertainties are negligible, except for 
$\bstodspipipi$ and $\bstodspi$, where we find deviations as large as 5\% when varying the central value of 
the constraints on the $\btodpipipi$ and $\btodpi$ reflections by $\pm$1 standard deviation.
We have checked our sensitivity to the signal model by varying the constraints on the width ratio and 
core Gaussian area fraction by one standard deviation (2\%).
We also include a systematic uncertainty of 1\% for 
neglecting the small radiative tail in the fit, which is estimated by comparing the
yields between our double Gaussian signal model and the sum of a Gaussian and Crystal Ball~\cite{cbal2} 
line shape. Taken together, we assign a 4\% uncertainty to 
the relative yields. For the $\Bsb$ branching fraction ratio, the total fitting uncertainty is 6.4\%.

Another difference between the $\xbtoxcpi$ and $\xbtoxcpipipi$ selection is the upper limit on the number of
tracks. The efficiencies of the lower track multiplicity requirements can be evaluated using the samples with higher track 
multiplicity requirements. Using this technique, we find corrections of $0.95\pm0.01$ for the $B^-$ and $\Lambda_b^0$ 
branching fraction ratios, and $0.99\pm0.01$ for the $\Bzb$ and $\Bsb$ branching fraction ratios.

We have also studied the PID efficiency uncertainty using a $D^{*+}$ calibration sample in data. Since
the PID requirements are either common to the signal and normalization modes, or in the case of the bachelor
pion(s), the selection is very loose, the uncertainty is small and we estimate a correction of $1.01\pm0.01$. We have
also considered possible background from $\xb\to\xc D_s^-$ which results in a correction of $0.99\pm0.01$.

All of our MC samples have a comparable number of events, from which we incur 3-4\% uncertainty in
the efficiency ratio determinations. 
The full set of systematic uncertainties and corrections are shown in Table~\ref{tab:syst}.
In total, the systematic uncertainty is $\sim$9\%, with correction factors that range from 1.01 to 1.07.

\begin{table*}[ht]
\begin{center}
\caption{Summary of corrections and systematic uncertainties to the ratio of branching fractions 
$\br(\xbtoxcpipipi)/\br(\xbtoxcpi)$.}
\begin{tabular}{lcccc}
\hline\hline
Source                        & \multicolumn{4}{c}{central value $\pm$ syst. error} \\
\hline
& \raisebox{-0.5ex}{$\Bzb$} & \raisebox{-0.5ex}{$B^-$} & \raisebox{-0.5ex}{$\Bsb$} & \raisebox{-0.5ex}{$\Lambda_b$} \\ [0.5ex]
\hline
Track Reconstruction   & \multicolumn{4}{c}{$1.00\pm0.04$} \\
IP/Vertex Resolution   & \multicolumn{4}{c}{$1.02\pm0.03$} \\
Production/Decay Model &  \multicolumn{4}{c}{$1.04\pm0.04$} \\
Best Cand. Selection     & $1.02\pm0.02$ & $1.01\pm0.01$ & $1.02\pm0.02$ & $1.03\pm0.02$ \\
Trigger Efficiency     & \multicolumn{4}{c}{$1.00\pm0.02$}  \\
Fitting                & $1.00\pm0.04$ & $1.00\pm0.04$ & $1.00\pm0.06$ & $1.00\pm0.04$  \\
Cut on \#Tracks     & $0.99\pm0.01$ & $0.95\pm0.01$ & $0.99\pm0.01$ & $0.95\pm0.01$ \\
PID                    & \multicolumn{4}{c}{$1.01\pm0.01$} \\
$\xc D_s^+$ background  & \multicolumn{4}{c}{$0.99\pm0.01$} \\
MC Statistics          & $1.00\pm0.04$ & ~~$1.00\pm0.03$~~ &~~ $1.00\pm0.04$~~ & $1.00\pm0.04$  \\
\hline
Total Correction        &  1.07 &  1.01 &   1.07 & 1.03  \\
Total Systematic (\%)   &  8.8   & 8.4   & 10.1 & 9.2       \\
\hline\hline
\end{tabular}
\label{tab:syst}
\end{center}
\end{table*}

\section{Results for {$\boldsymbol{\xbtoxcpipipi}$} }

The results for the ratios of branching ratios are
\begin{align}
{\br(\btodpipipi)\over\br(\btodpi)} = 2.38\pm0.11\pm0.21 \nonumber \\
{\br(\btodzeropipipi)\over\br(\btodzeropi)} = 1.27\pm0.06\pm0.11 \label{eq:cfbf} \\
{\br(\bstodspipipi)\over\br(\bstodspi)} = 2.01\pm0.37\pm0.20 \nonumber \\
{\br(\LbtoLcpipipi)\over\br(\LbtoLcpi)} = 1.43\pm0.16\pm0.13, \nonumber 
\end{align}

\noindent where the first uncertainty is statistical and the second is systematic.
These measurements are all substantially more precise than the current world average values. 
Naively, one might have expected the four branching fraction ratios to
be nearly equal. The observed differences may be explained in terms of the contributing Feynman diagrams. From
Fig.~\ref{fig:feyn}, we see that the primary contribution to $\Bzb\to D^+\pi^-(\pi^+\pi^-)$ and $\Bsb\to D_s^+\pi^-(\pi^+\pi^-)$ is
from a single decay diagram, an external tree diagram. On the other hand the
$\B^-\to\D^0\pi^-(\pi^+\pi^-)$ and $\Lb\to\Lc\pi^-(\pi^+\pi^-)$ amplitudes receive contributions from both 
external and color-suppressed tree diagrams. This would suggest that the interference tends to be more 
constructive in $\btodzeropi$ and $\LbtoLcpi$ than in $\btodzeropipipi$ and $\LbtoLcpipipi$ 
respectively. The role of the various contributing topological amplitudes and the strong phases in
$B\to D\pi$ is discussed in the literature~\cite{rosner}.
In general we see the branching fractions for the $\xc\pi\pi\pi$ final states are at least as 
large or even twice as large as the single-$\pi$ bachelor states.

\section{Kinematic Distributions and Mass Spectra in the {$\boldsymbol{\pi^-\pi^+\pi^-}$} System}
\label{sec:kin}

\begin{figure}[ht]
\centering
\includegraphics[width=125mm]{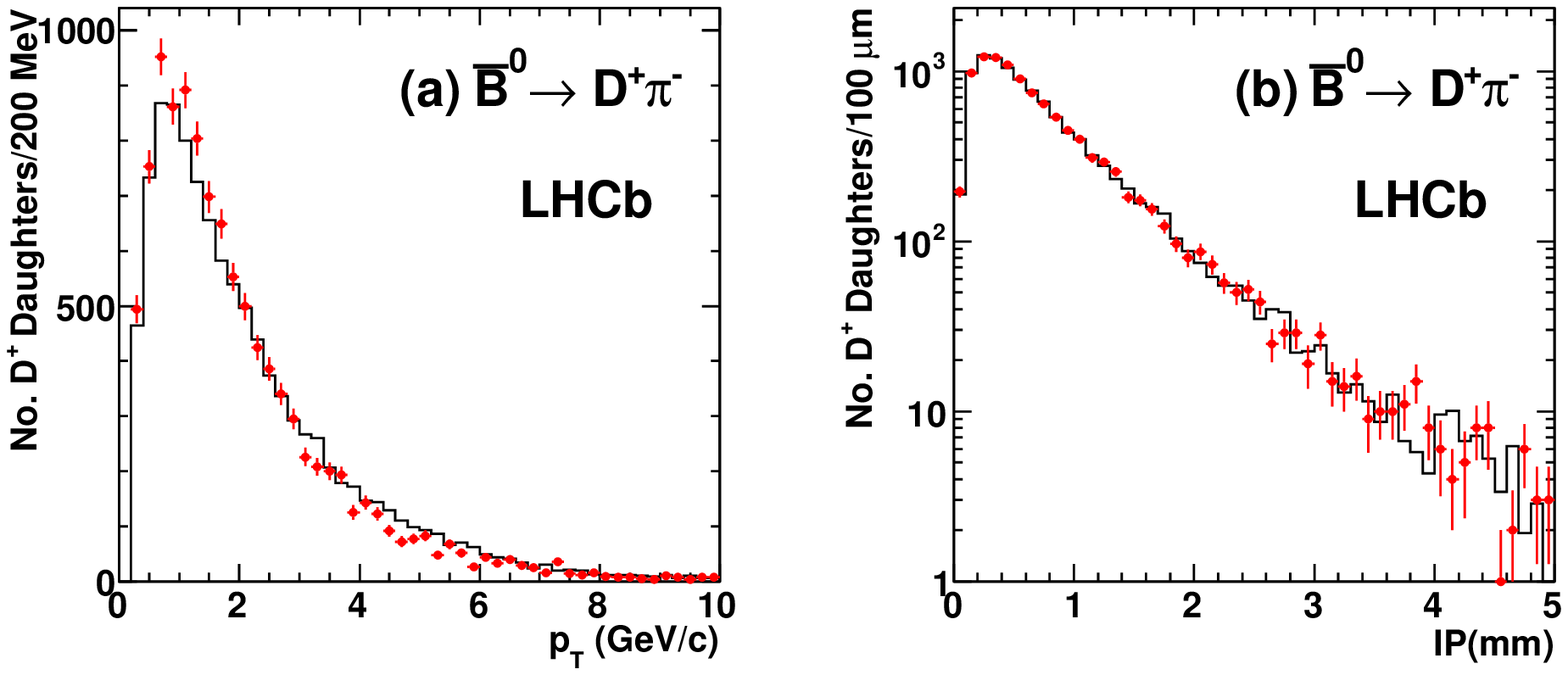}
\includegraphics[width=125mm]{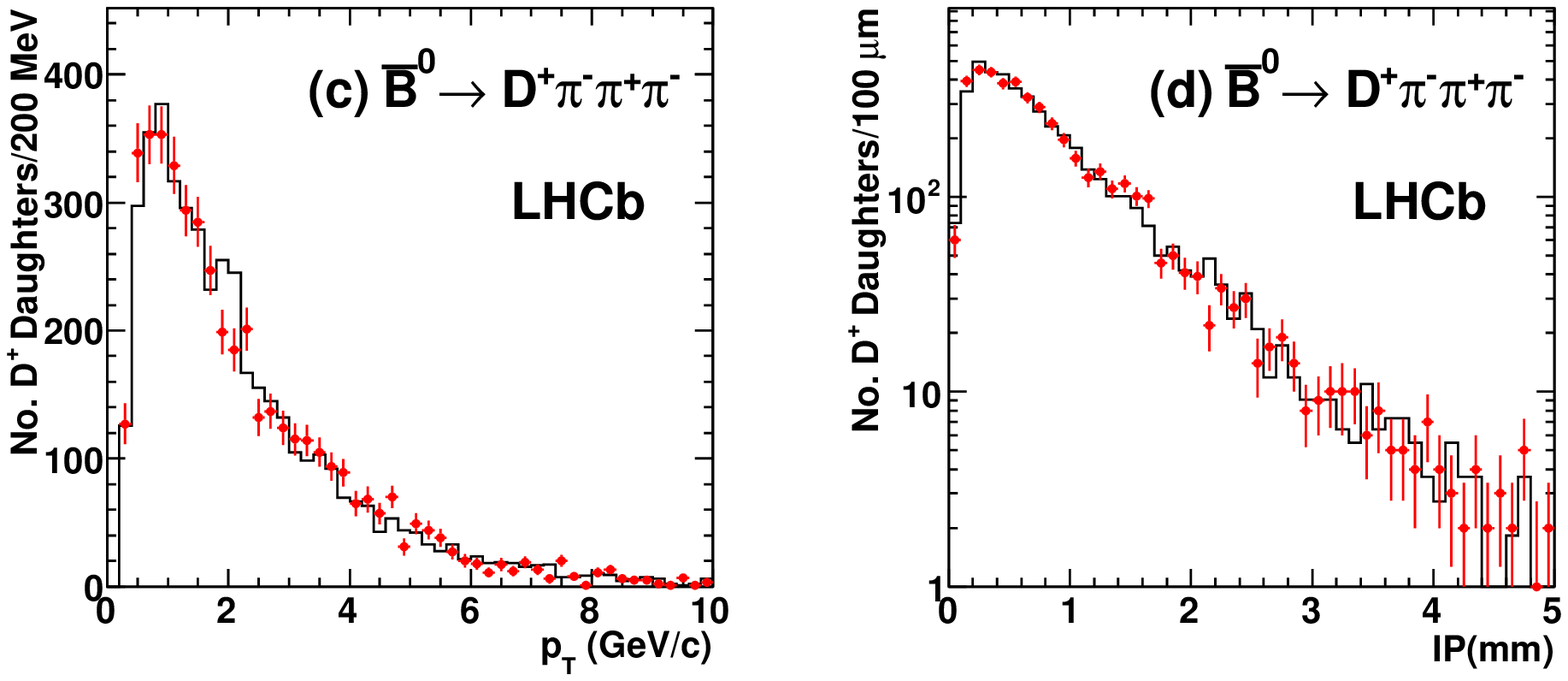}
\caption{Comparisons of the $p_T$ and IP spectra for the daughters from the $D^+$ in
$\btodpi$ [(a) and (b)], and from the $D^+$ in $\btodpipipi$ [(c) and (d)]; 
Points with error bars are data and the solid lines are simulation.}
\label{fig:kindistD}
\end{figure}

\begin{figure}[ht]
\centering
\includegraphics[width=125mm]{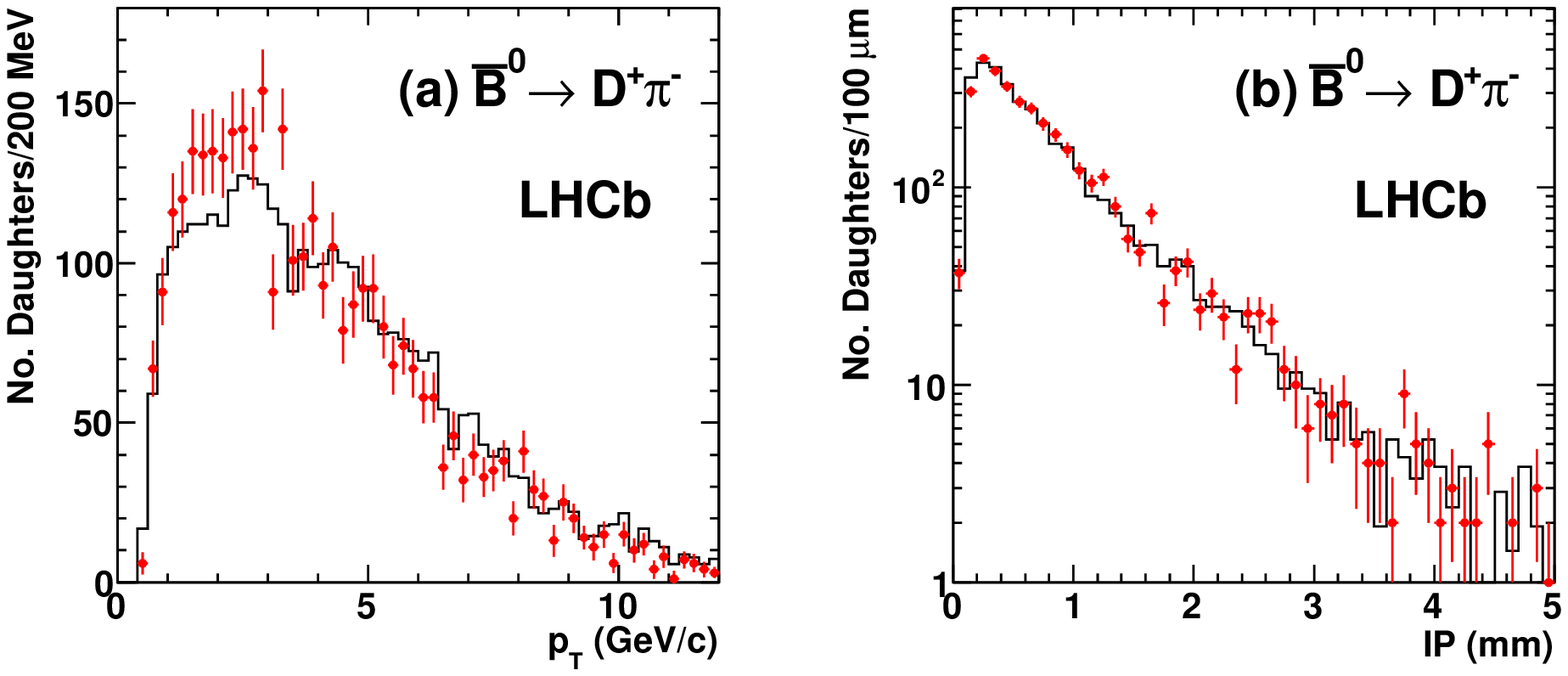}
\includegraphics[width=125mm]{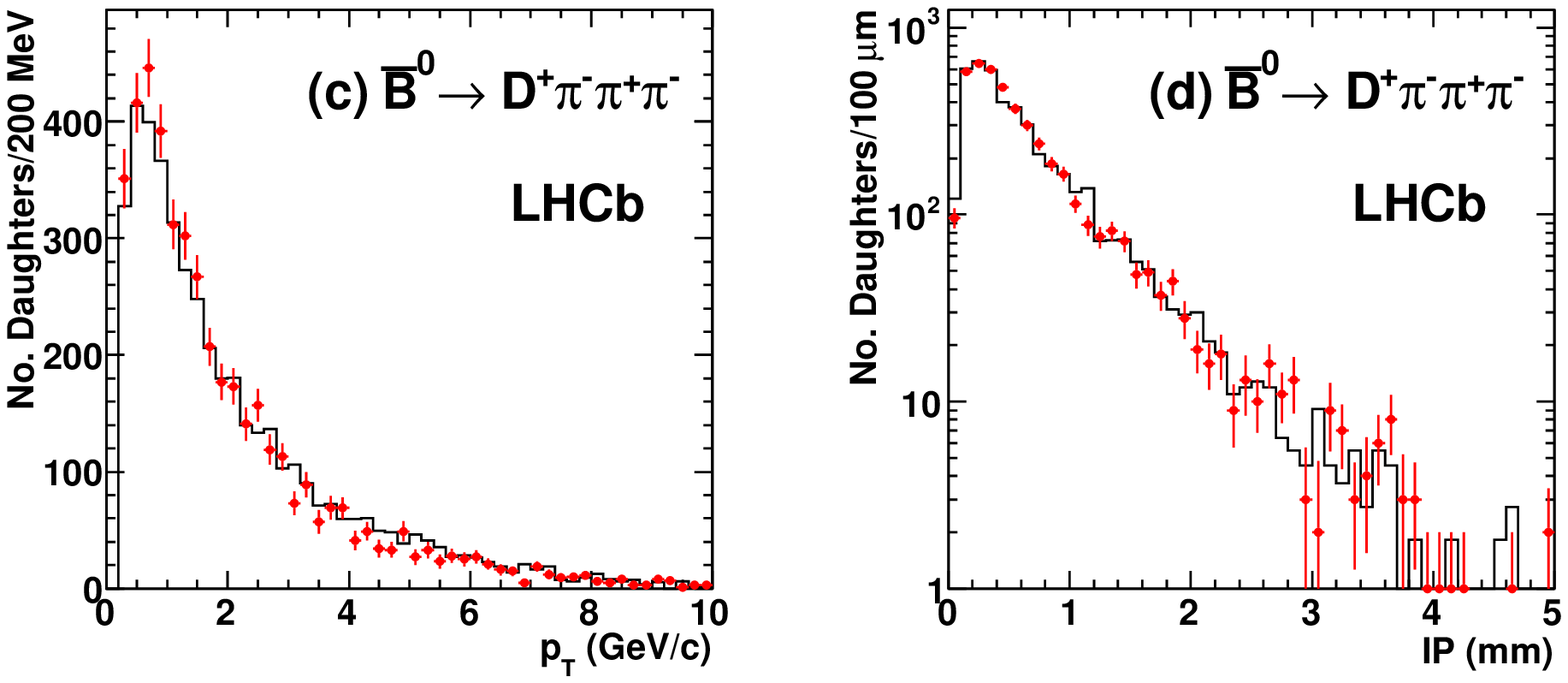}
\caption{Comparisons of the $p_T$ and IP spectra for the bachelor pion in 
$\btodpi$ [(a) and (b)], and for the 3 pions in $\btodpipipi$ [(c) and (d)].
Points with error bars are data and the solid lines are simulation.}
\label{fig:kindistPi}
\end{figure}

Since we rely on MC simulation to estimate signal efficiencies, we now compare
a few distributions between signal MC simulation and data. The higher
signal yield $\btodpi$ and $\btodpipipi$ decay modes are used, and for each we perform a sideband
subtraction, where the signal region includes candidates within 50~MeV/$c^2$ of
the $B^0$ mass, ($m_{\Bz}$)~\cite{pdg}, and the sidebands $60<|M-m_{\Bz}|<110$~MeV/$c^2$.
For both data and simulation, we require events to pass any L0 trigger, and signal 
candidates must satisfy the HLT1 and HLT2 triggers described in Sec.~\ref{sec:det}. Clearly, two of the
most important quantities used in our candidate selection are the $p_T$ and IP of the daughters from
the $D^+$ and the recoiling pion(s). Figure~\ref{fig:kindistD} compares
the $p_T$ and IP distributions of the $D^+$ daughters in data to those
from signal MC simulation. Figure~\ref{fig:kindistPi}
shows the corresponding comparisons for the recoiling pion(s) in the respective $B$ decay.
Overall, the agreement between data and MC simulation is very good. 

\begin{figure}[ht]
\centering
\includegraphics[width=75mm]{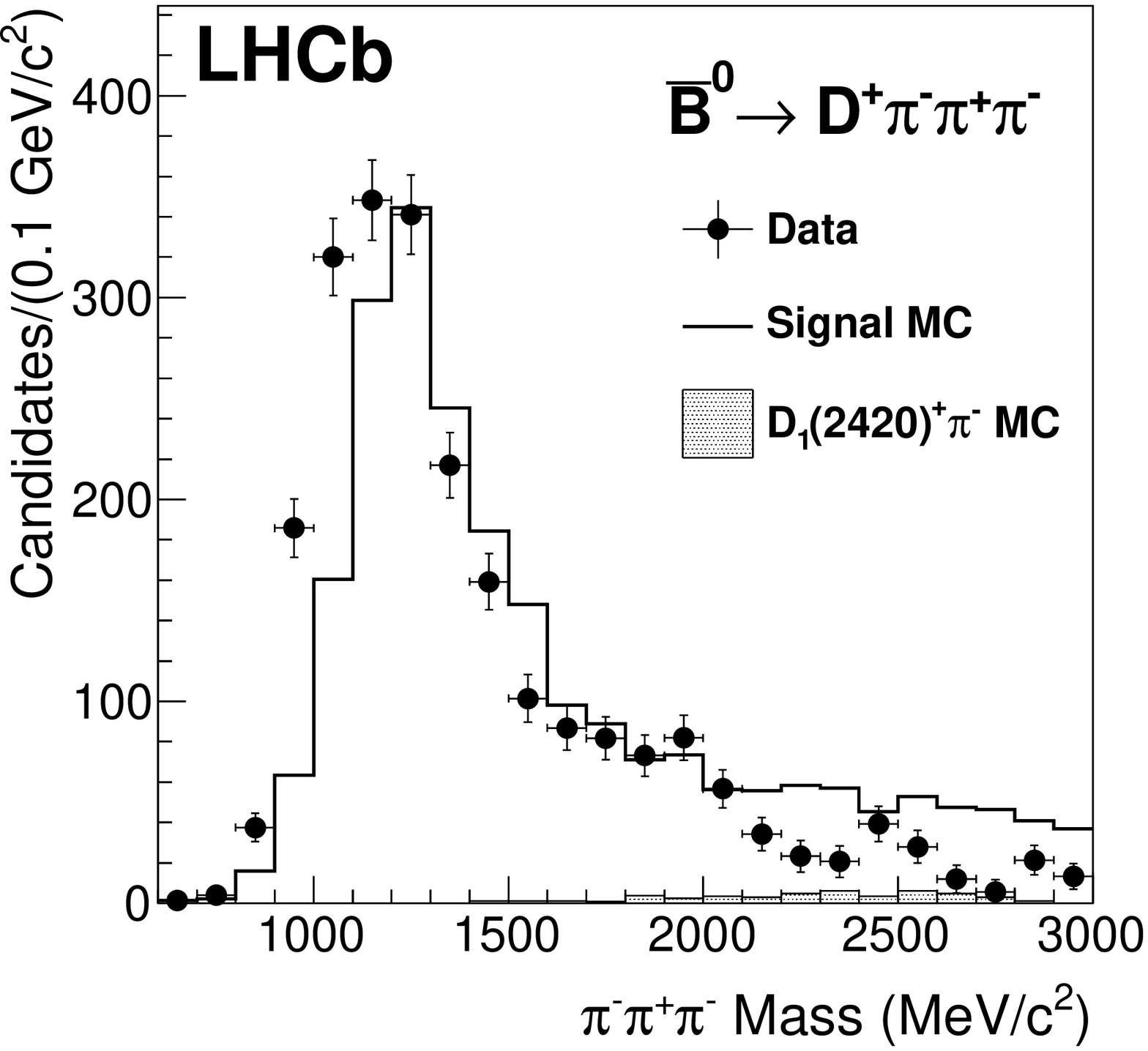}
\includegraphics[width=75mm]{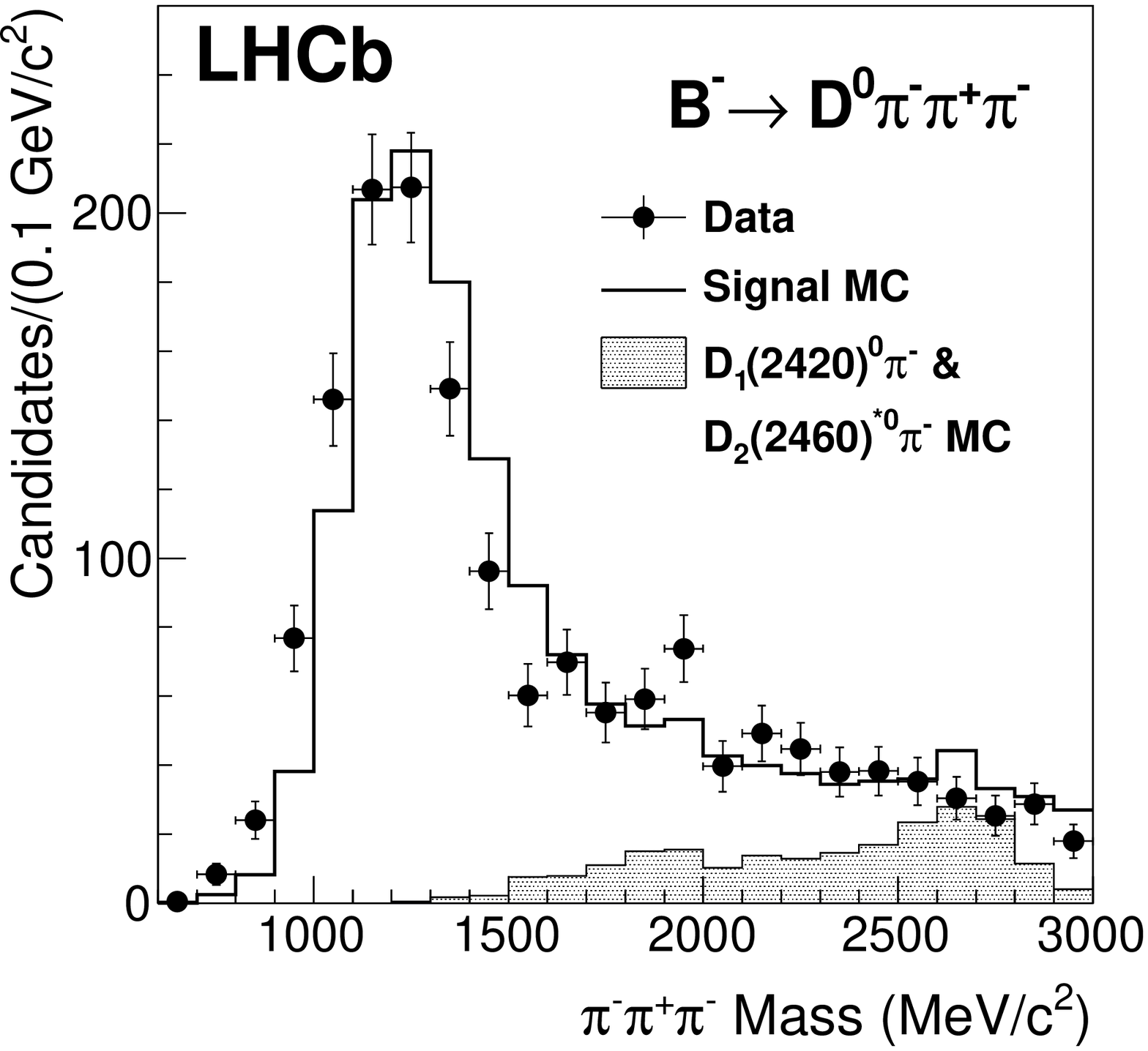}
\includegraphics[width=75mm]{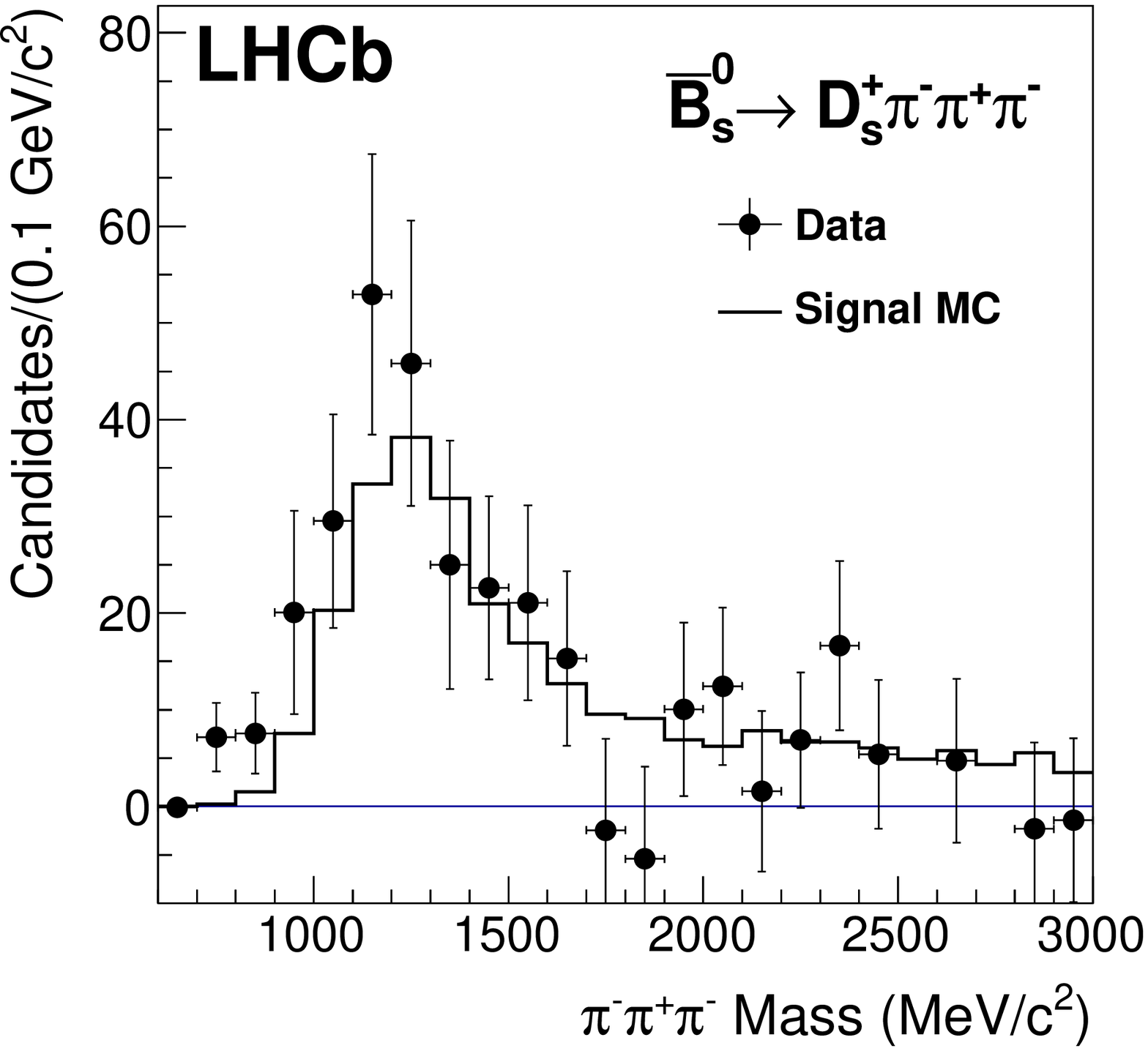}
\includegraphics[width=75mm]{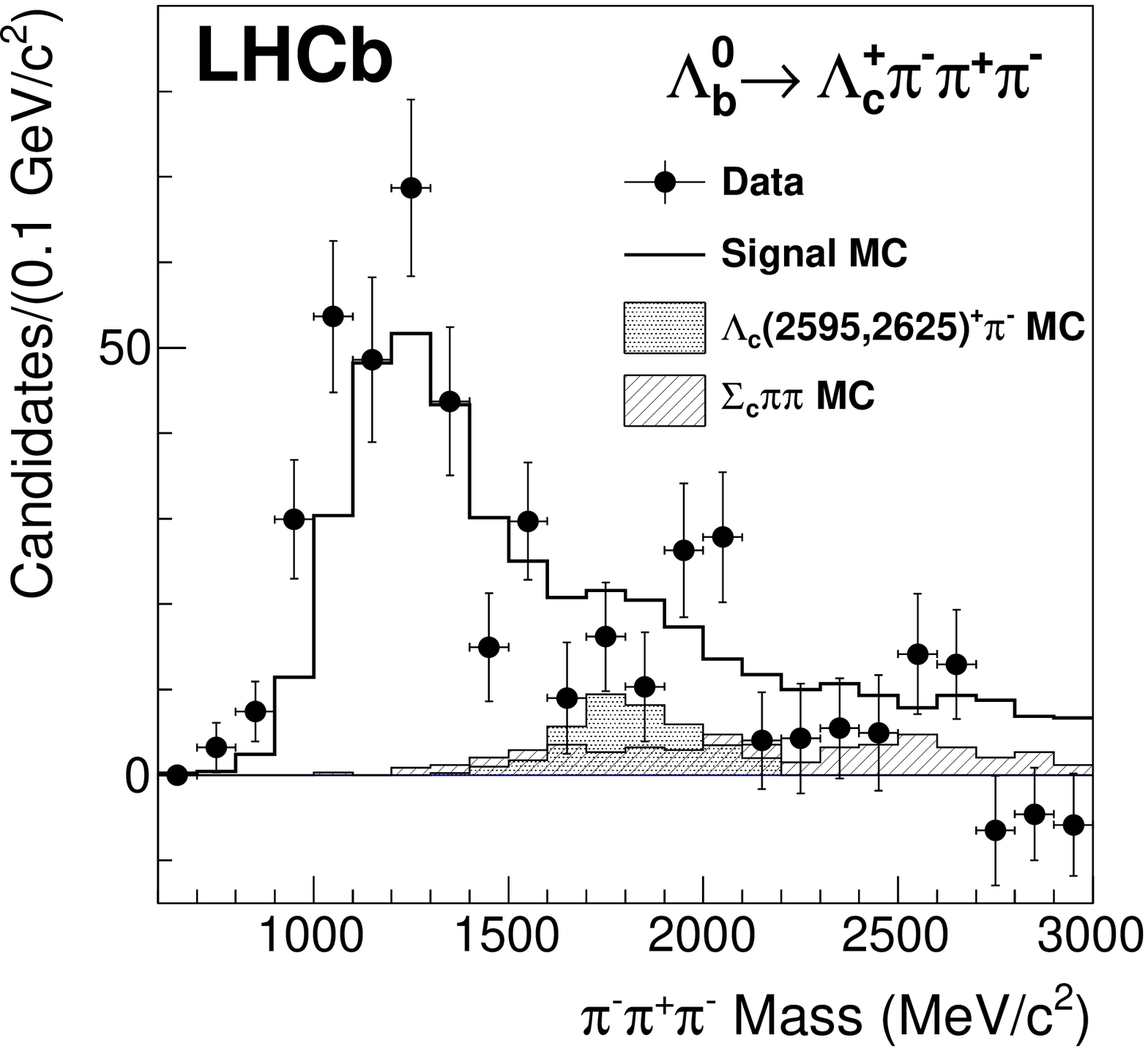}
\caption{Invariant mass of the $3\pi$ system in $\btodpipipi$ (top left), 
$\btodzeropipipi$ (top right), $\bstodspipipi$ (bottom left) and 
$\LbtoLcpipipi$ (bottom right) decays. 
The data are the points with error bars and the simulations are the solid lines and shaded regions.}
\label{fig:a1mass}
\end{figure}

It is also interesting to examine the $\pi^-\pi^+\pi^-$ invariant mass spectra for 
the four signal decay modes. Here, we use the sPlot method~\cite{sPlot} to obtain the
underlying signal spectra, based on the event-by-event $b$-hadron mass signal and background probabilities.
The $\pi^-\pi^+\pi^-$ mass spectra are shown in Fig.~\ref{fig:a1mass}, along
with signal MC shapes that are normalized to the same yield as data. We also show several
resonant contributions: $D_1(2420)^+$ (2\%), $D_1(2420)^0$ and $D_2^*(2460)^0$ (14\% in total),
$\Lambda_c(2595)^+$ and $\Lambda_c(2625)^+$ (9\% total), and $\Sigma_c^0$ and $\Sigma_c^{++}$ (12\% total),
where the quantities in parentheses are the normalizations relative to the total (see Sec.~\ref{sec:excCharm}.)
A prominent structure at low mass, consistent with the $a_1(1260)^-$ is evident for all decay modes, along with a long tail 
extending to 3~GeV/$c^2$. 
In all cases, the 3$\pi$ mass spectrum appears shifted toward lower mass as compared to the 
MC simulation. The simulated value for the $a_1(1260)^-$ mass is 1230~MeV/$c^2$, which is equal to the 
central value given in the PDG~\cite{pdg} of $(1230\pm40)$~MeV/$c^2$. Besides having a large uncertainty, 
the mass as obtained by experiment may be
process-dependent, so it is difficult to draw any definitive conclusion from this
shift. Since both the reconstruction and trigger efficiency are flat through this mass region,
this small shift in mass does not introduce any significant systematic uncertainty in the branching
fraction measurement.

\begin{figure}[ht]
\centering
\includegraphics[width=150mm]{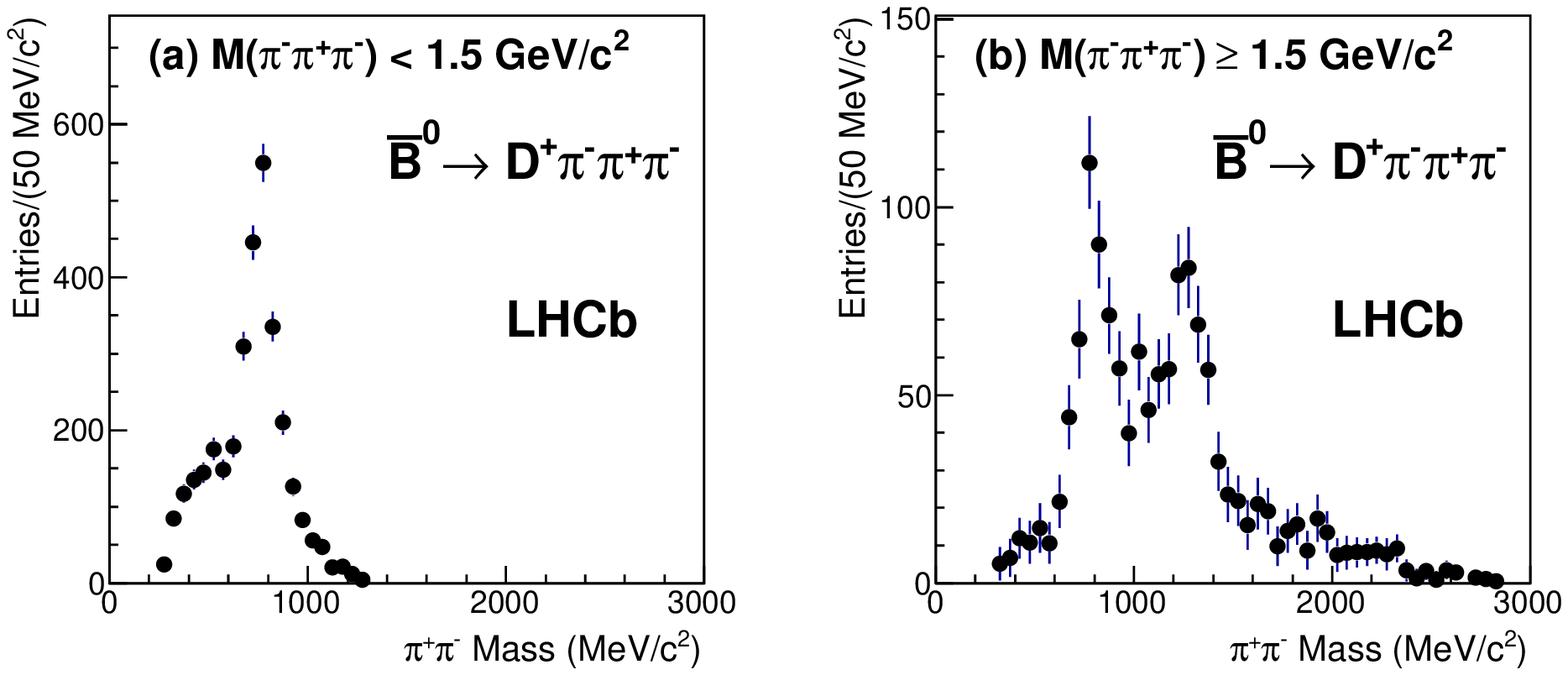}
\includegraphics[width=150mm]{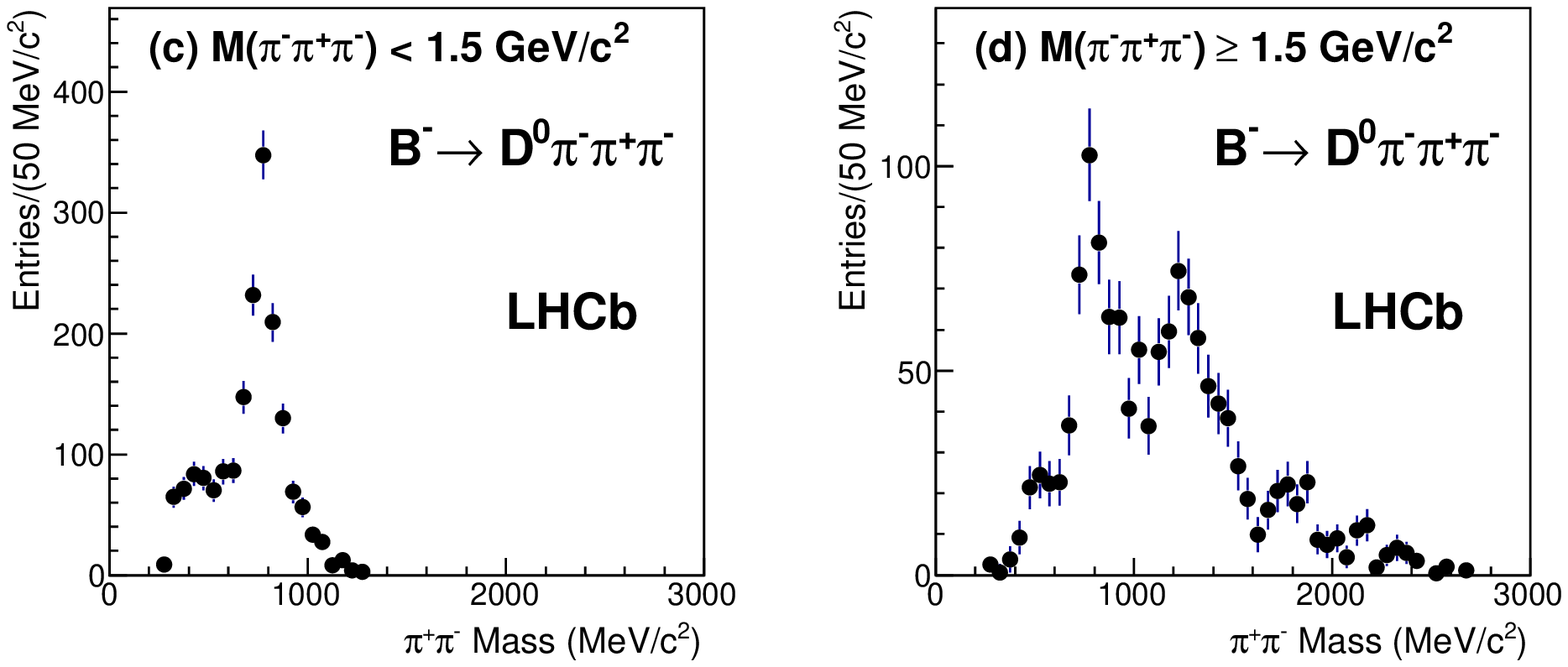}
\caption{$\pi^+\pi^-$ invariant mass (2 combinations per $\bzb$ candidate) in the $3\pi$ system for $\btodpipipi$ 
when (a) $M(\pi^-\pi^+\pi^-)<1.5$~GeV/$c^2$ and (b) $M(\pi^-\pi^+\pi^-)\ge 1.5$~GeV/$c^2$. The
corresponding plots for $\btodzeropipipi$ are shown in (c) and (d).}
\label{fig:pipiResInB0}
\end{figure}

We have also looked at the di-pion invariant masses within the 3$\pi$ system, shown for $\btodpipipi$(a,b)
and $\btodzeropipipi$(c,d) in Fig.~\ref{fig:pipiResInB0}.
Contributions from the narrow excited charm states, which are discussed in 
Sec.~\ref{sec:excCharm}, are excluded. 
In all cases, in the low $M(\pi^-\pi^+\pi^-)$ mass region, we see a dominant $\rho^0\pi^-$ contribution, consistent
with the $a_1(1260)^-$ resonance. In the 
higher  $M(\pi\pi\pi)$ regions there appears to be an additional resonant structure, consistent 
with the $f_2(1270)$ state, in addition
to the $\rho^0$ contribution. Similar spectra are found for $\bstodspipipi$ and
$\LbtoLcpipipi$ (not shown.) The $f_2(1270)$ has been previously seen in 
$\btodstarpipipi$~\cite{b2dstarf2pi}.
The like-sign di-pion invariant mass spectra do not show any resonant features. 

\section{Contributions from Excited Charm Hadrons}
\label{sec:excCharm}

Within the $\xbtoxcpipipi$ final state, we search for $D_1(2420)$, $D_2^*(2460)$, 
$\Lambda_c(2595)^+$, $\Lambda_c(2625)^+$ and $\Sc^{0,++}$, which 
may decay to $D$ or $\Lc$ with an accompanying $\pi^{\pm}$ or $\pi\pi$ pair.
To search for $\xc^*\to\xc\pi^+\pi^-$ intermediate states, we select events in the $b$-hadron signal region 
($\pm60$~MeV/$c^2$ around the nominal mass) 
and compute the invariant mass difference 
$\Delta M_{\pi\pi}\equiv M(\xc\pi^+\pi^-)-M(\xc)$ 
(two combinations per $b$-hadron candidate.) 
For the $\Lb\to\Sc^{0,++}\pi^{\pm}\pi^-,~\Sc^{0,++}\to\Lc\pi^{\pm}$, we use 
$\Delta M_{\pi}\equiv M(\xc\pi^{\pm})-M(\xc)$ in a similar way
(one (two) $\Sc^{++}$ ($\Sc^0$) candidates per $\Lb$ decay.)
We also have looked in the upper mass sidebands, and the $\Delta M_{\pi\pi}$ and $\Delta M_{\pi}$
distributions are consistent with a smooth background shape with no signal component.
We look at all data, irrespective of trigger, to establish signal
significances, but for the branching fraction measurement, we use the same
trigger requirements described in Sec.~\ref{sec:kin}. We choose only
one candidate per event using the same criteria as discussed previously. We normalize
the rates to the respective inclusive $\xbtoxcpipipi$ decay, using the same trigger selection
as above. We show only the $\Delta M_{\pi\pi}$ and $\Delta M_{\pi}$ distributions after the specified 
trigger, since the distributions before the trigger are quite similar, except they typically have 
25$-$30\% larger yields than the ones shown.

\begin{figure}[ht]
\centering
\includegraphics[width=75mm]{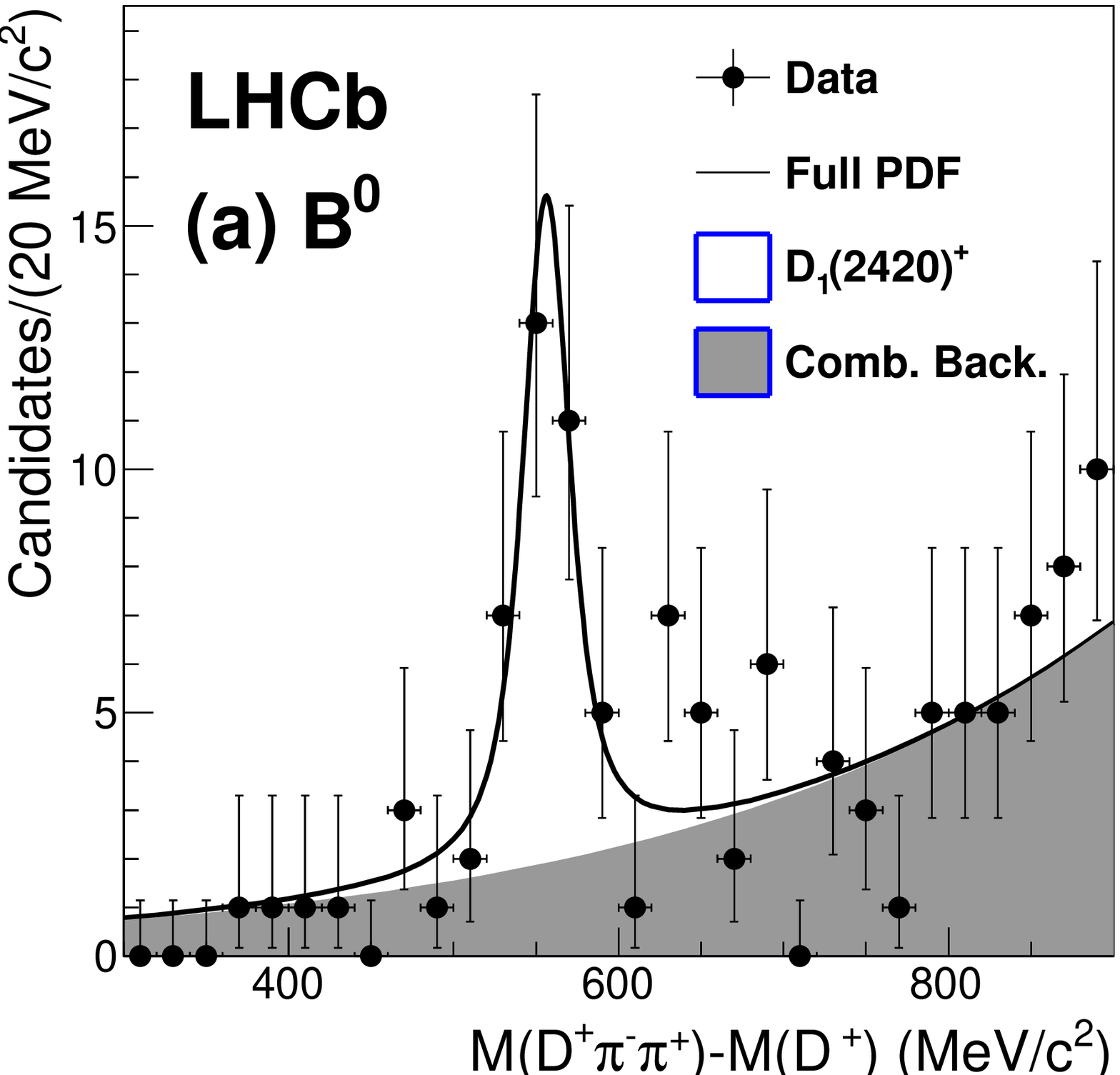}
\includegraphics[width=75mm]{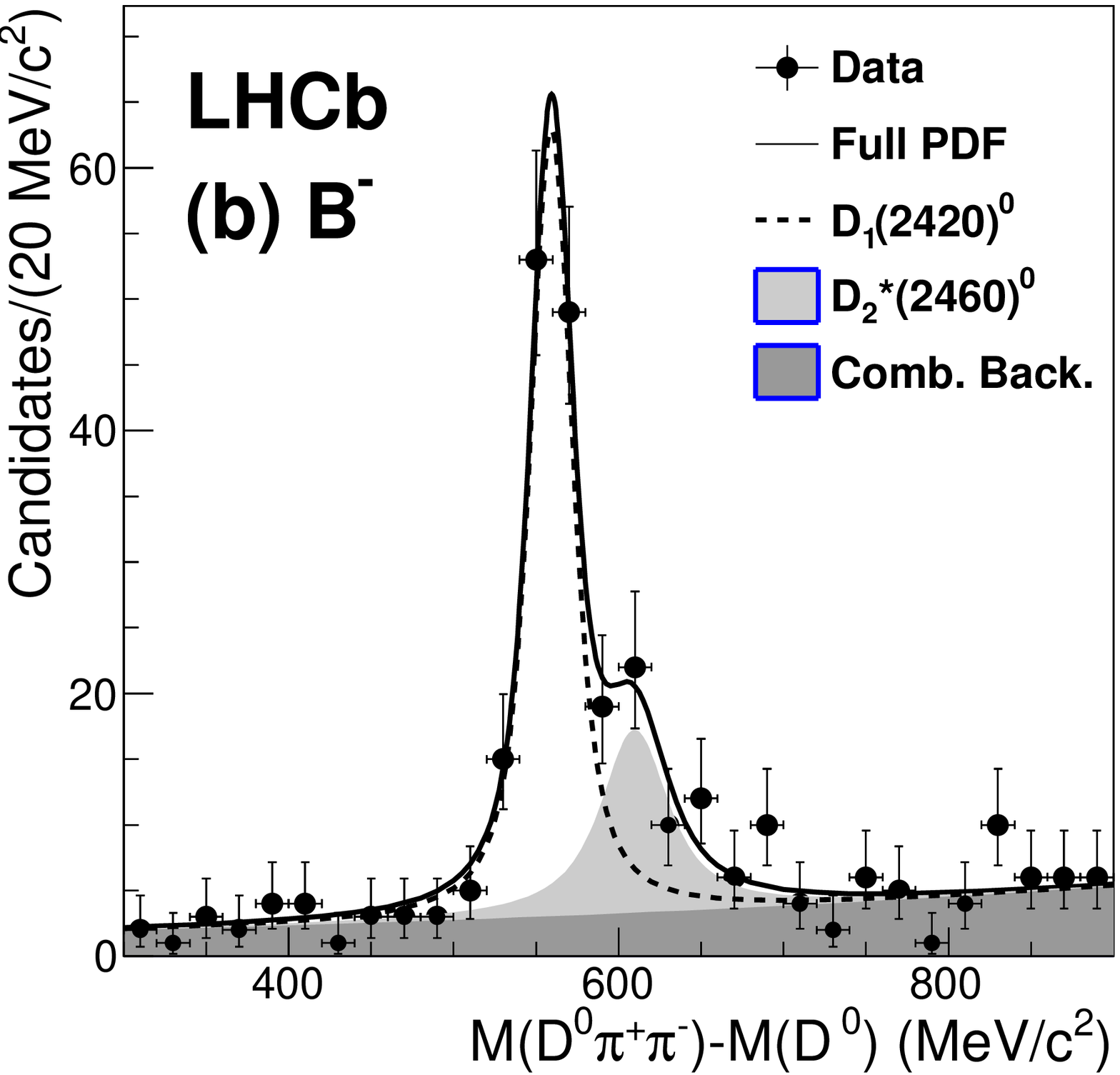}
\includegraphics[width=75mm]{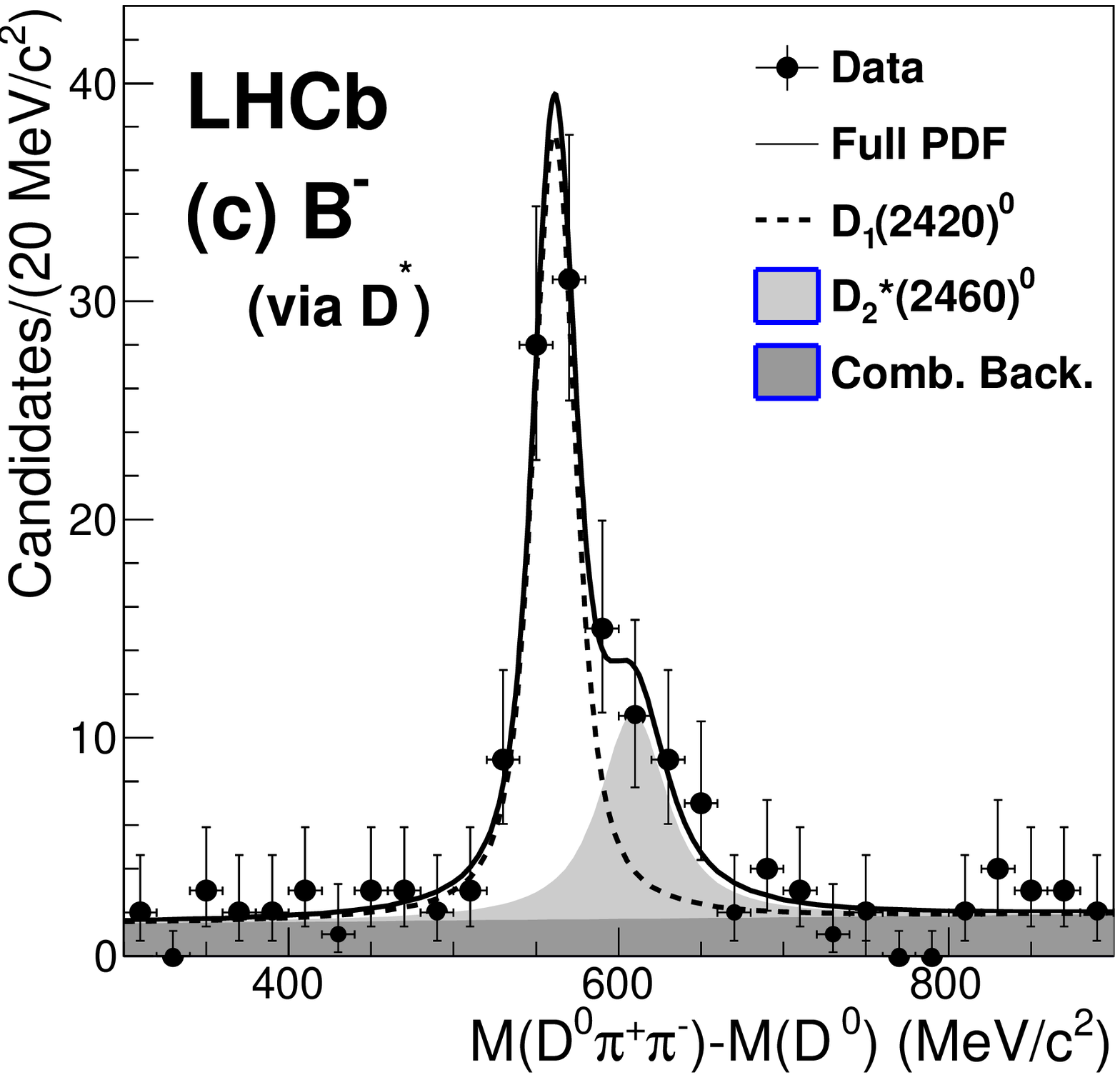}
\includegraphics[width=75mm]{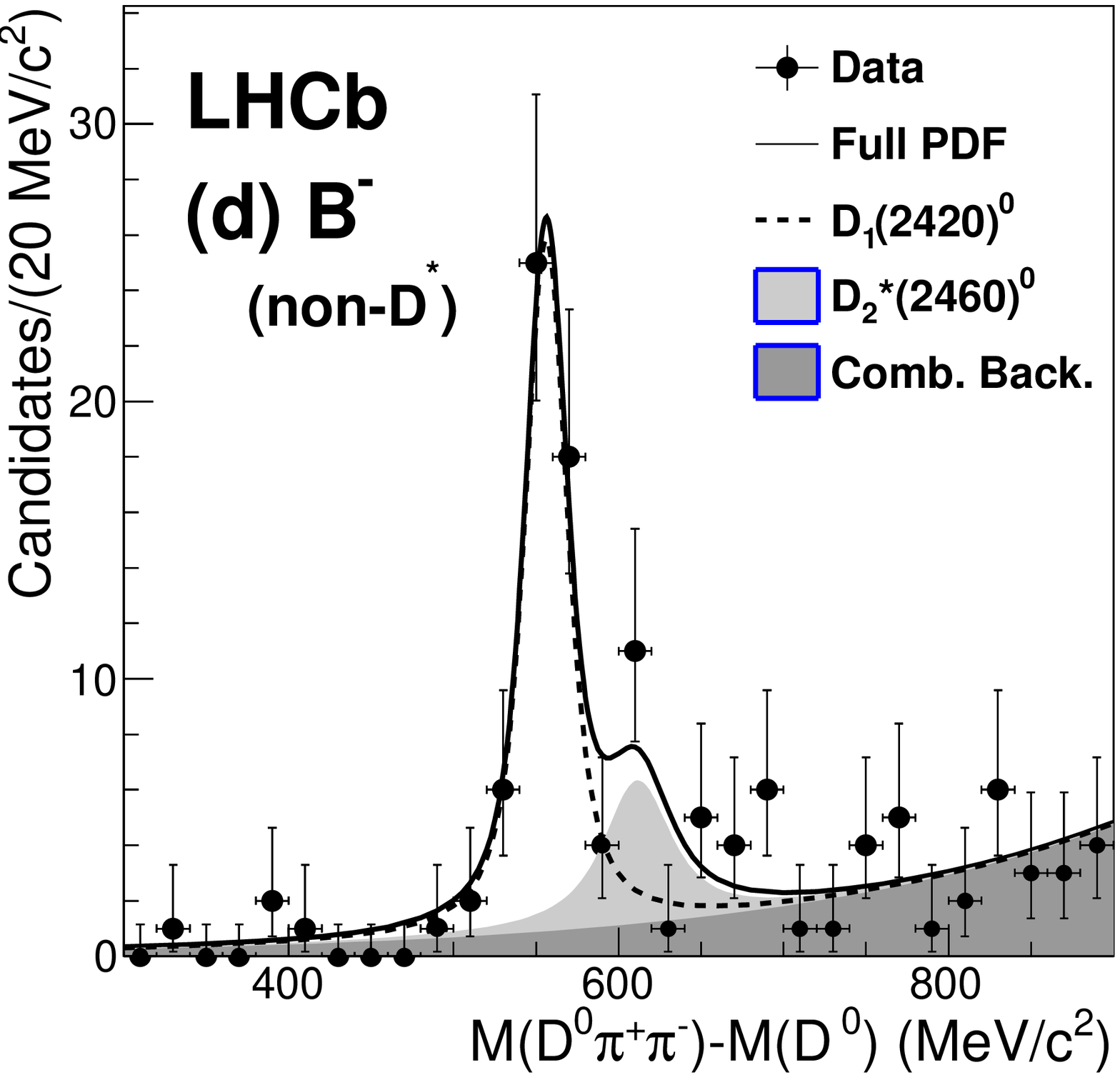}
\caption{Invariant mass difference $M(D\pi^-\pi^+) - M(D)$, for 
(a) $\btodpipipi$ signal candidates, (b) $\btodzeropipipi$ signal candidates, 
(c) $\btodzeropipipi$ through a $D^{*+}$ intermediate state, and
(d) $\btodzeropipipi$ not through a $D^{*+}$ intermediate state.
The signal components are the white region (and lightly shaded regions for $\btodzeropipipi$), 
and the background component is the darker shaded region.}
\label{fig:B02DA1_ExcCharm}
\end{figure}

\begin{figure}[ht]
\centering
\includegraphics[width=75mm]{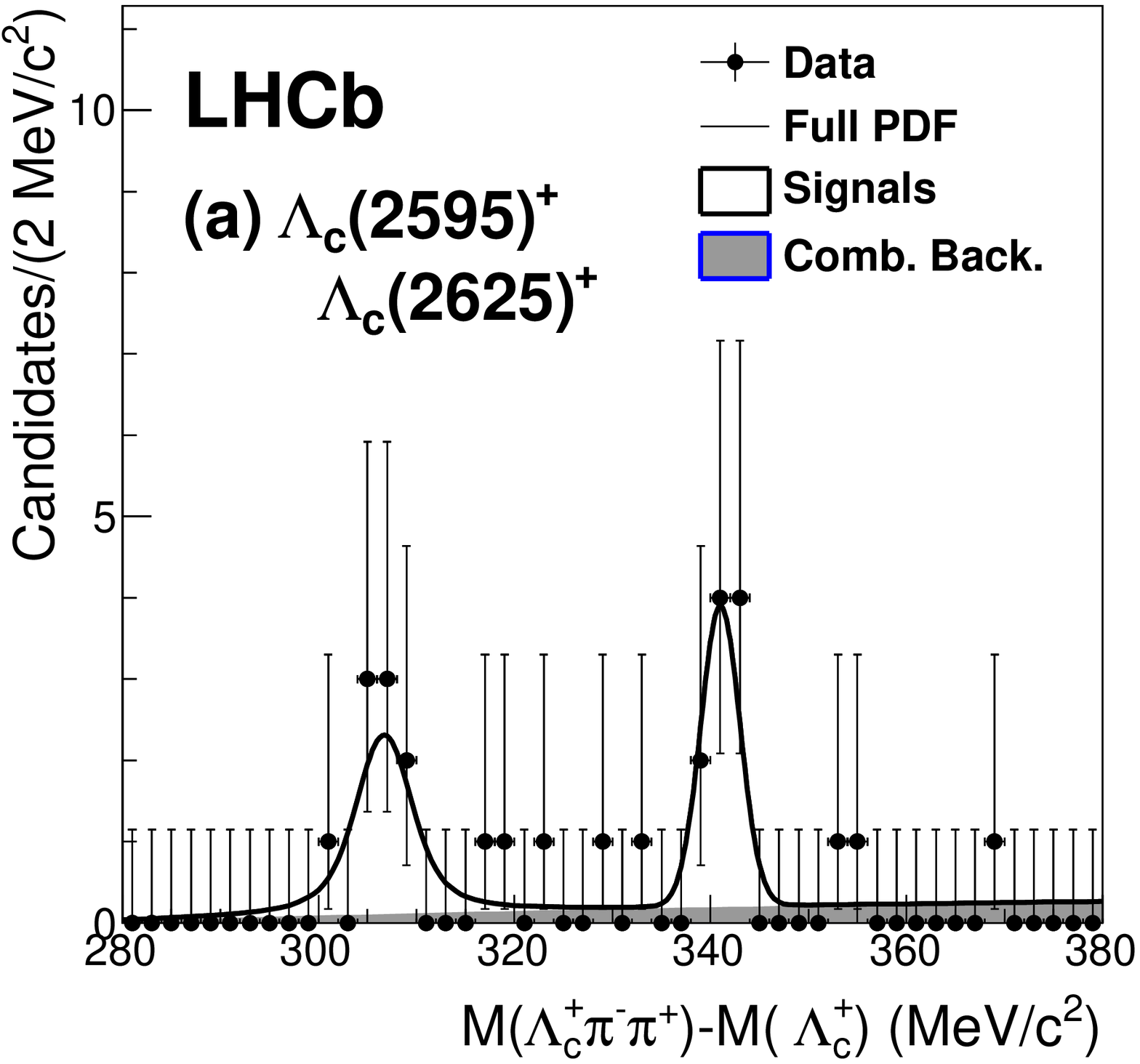}
\includegraphics[width=75mm]{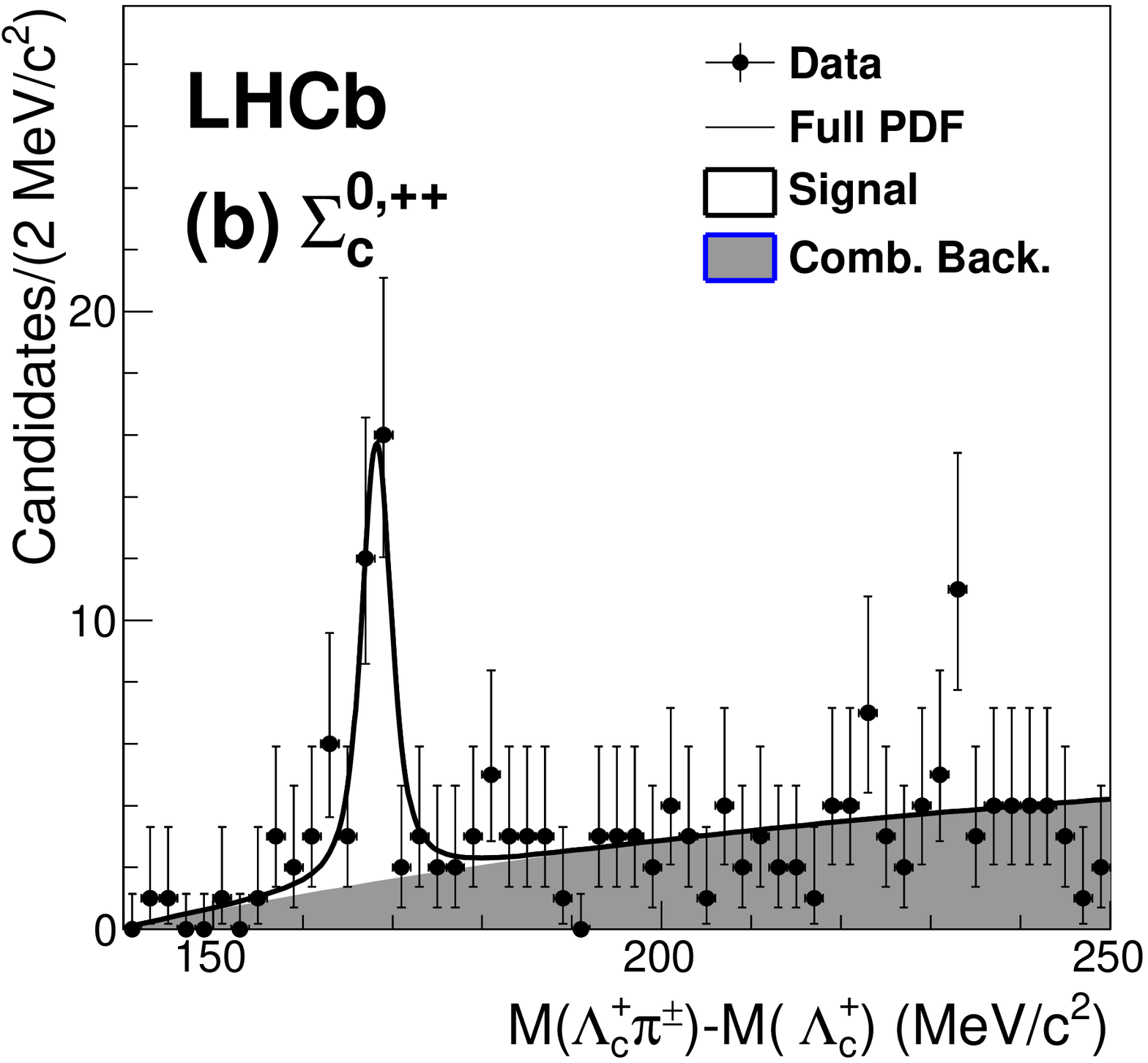}
\includegraphics[width=75mm]{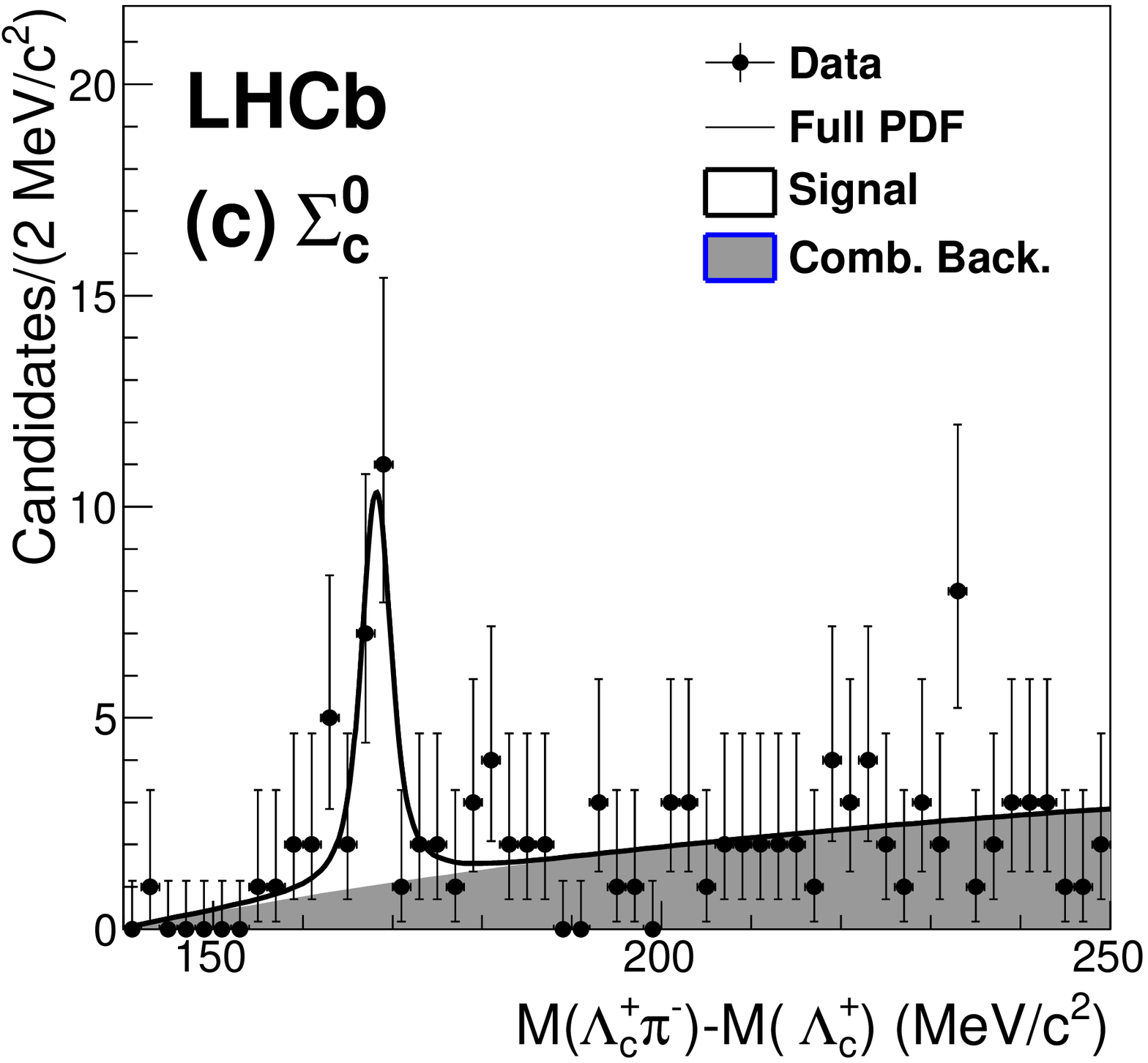}
\includegraphics[width=75mm]{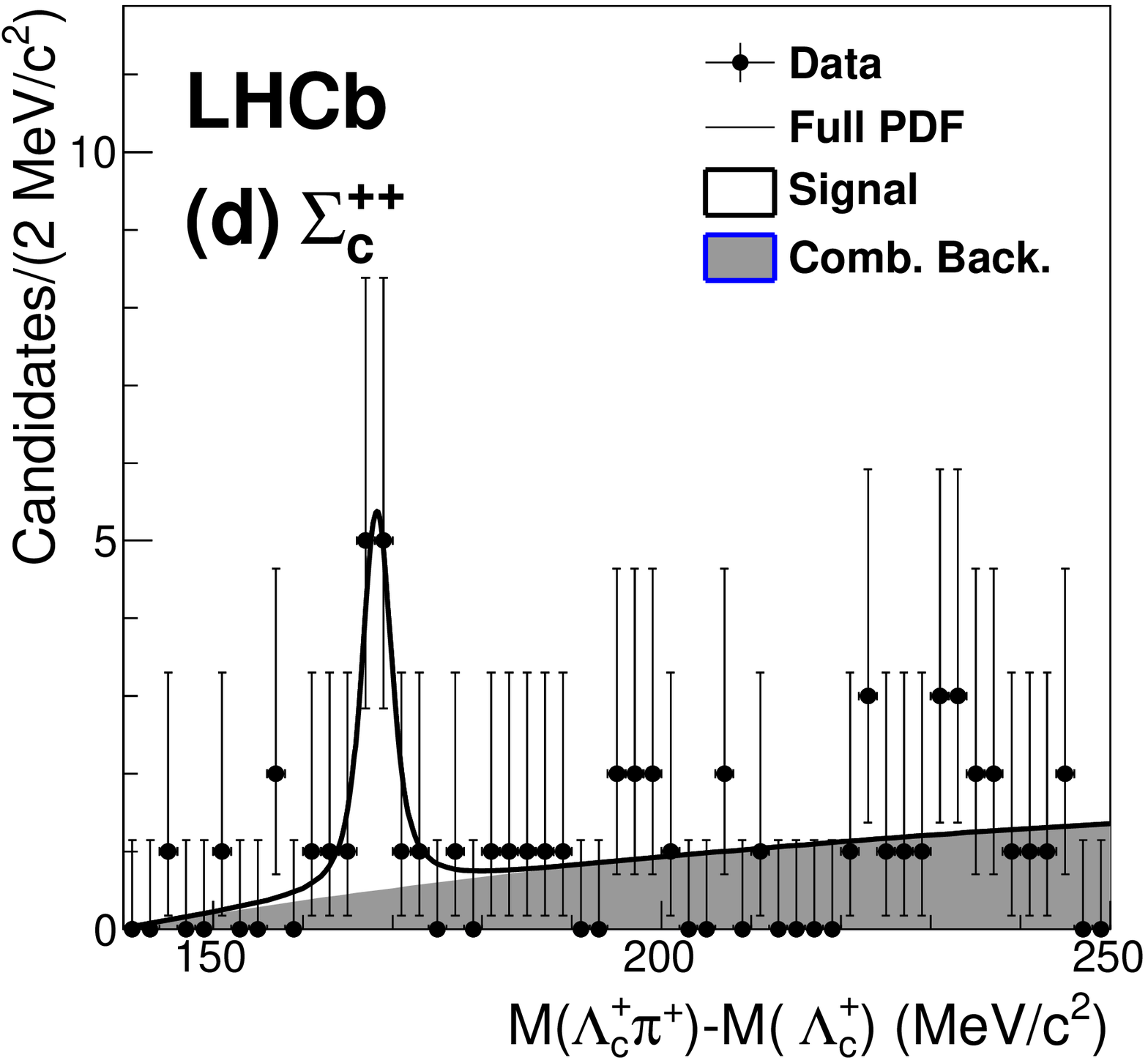}
\caption{Intermediate resonances contributing to the $\LbtoLcpipipi$ decay. Shown
are distributions for (a) $M(\Lc\pi^-\pi^+) - M(\Lc)$, with $\Lambda_c(2595)^+$ and
$\Lambda_c(2625)^+$ contributions, (b) $M(\Lc\pi^{\pm}) - M(\Lc)$ (3 combinations per $\Lb$ candidate) 
(c)  $M(\Lc\pi^{-}) - M(\Lc)$ (2 combinations per $\Lb$ candidate), and (d) $M(\Lc\pi^{+}) - M(\Lc)$ 
(1 combination per $\Lb$ candidate), showing the intermediate $\Sc$ states. The line is a fit
as described in the text, and the shaded region is the fitted background.}
\label{fig:Lb2LcA1_ExcCharm}
\end{figure}

The $\Delta M_{\pi\pi}$ distributions for $\bzb$ and $\B^+$ are shown in 
Fig.~\ref{fig:B02DA1_ExcCharm} and the $\Delta M_{\pi}$ for $\Lambda_b^0$
are shown in Fig. ~\ref{fig:Lb2LcA1_ExcCharm}.
For $B_s^0$, the size of the data sample is insufficient to observe the excited $D_s$ states in these hadronic decays.

Signal yields are determined using unbinned extended maximum likelihood fits.
Starting with $\bzb$ (Fig.~\ref{fig:B02DA1_ExcCharm}(a)), we see an excess at $\Delta M_{\pi\pi}\sim560$ MeV/$c^2$, 
consistent with the $D_1(2420)^+$.
We fit the distribution to the sum of a signal Breit-Wigner shape convoluted with a
Gaussian resolution, and an exponential background shape. The full width is fixed to 25~MeV/$c^2$~\cite{pdg}
and the mass resolution is set to 7.5~MeV/$c^2$ based on simulation. 
The fitted yield is $33\pm8$ events and the fitted mean is $(562\pm4)$~MeV/$c^2$,
consistent with the expected value. If the width is allowed to float, we find $(22.7\pm8.0({\rm stat}))$~MeV/$c^2$, 
also in agreement with the world average.
Prior to applying the specific trigger selection, we find $40\pm9$ signal events, 
corresponding to a statistical significance of
6.8 standard deviations (for one degree of freedom) as determined from the difference in log-likelihoods, 
$\sqrt{-2\Delta LL}$, where the difference is taken between the signal yield taken
as a free parameter and fixed to zero. 

The $\Delta M_{\pi\pi}$ distributions for $B^-$ displayed in Fig.~\ref{fig:B02DA1_ExcCharm}(b) show not only the
$D_1(2420)^0$, but also a shoulder at $\sim$600~MeV/$c^2$, consistent with the $D_2^*(2460)^0$. Hence, we allow
for both $D_1(2420)^0$ and $D_2^*(2460)^0$ signal components, and fix their full widths to the
PDG values~\cite{pdg} of 20.4~MeV/$c^2$ and 42.9~MeV/$c^2$, respectively. The means and yields are left as 
free parameters in the fit. The fitted $D_1(2420)^0$ and $D_2^*(2460)^0$ yields are  $124\pm14$ and $49\pm12$,
with masses that are consistent with the expected values. The respective signal yields before the trigger requirement
are $165\pm17$ and $63\pm15$ events, with corresponding statistical significances of 10.5 and 5.5 standard
deviations for the $D_1(2420)^0$ and $D_2^*(2460)^0$, respectively.
These $B^0$ and $B^-$ decays have also been observed by Belle~\cite{belleb0tod2420pi}.

We have also measured the relative fractions of $D_1(2420)^0$ and $D_2^*(2460)^0$ that do
or do not decay through $D^{*+}$ by taking the subset of candidates with 
$M(D^0\pi^+)-M(D^0)\le 150$~MeV/$c^2$ or $M(D^0\pi^+)-M(D^0)>150$~MeV/$c^2$, respectively. The
corresponding $\Delta M_{\pi\pi}$ distributions are shown in Fig.~\ref{fig:B02DA1_ExcCharm}(c) and
Fig.~\ref{fig:B02DA1_ExcCharm}(d).
A fit is made to the data as discussed previously, and the yields are summarized in Table~\ref{tab:excCharm_yields}.

For $\Lambda_b^0$ (see Fig.~\ref{fig:Lb2LcA1_ExcCharm}(a)), we find two well-separated peaks in 
the $\Delta M_{\pi\pi}$ distribution, one at $\sim$307~MeV/$c^2$,
and a second at $\sim$340~MeV/$c^2$, consistent with the expected values for the $\Lambda_c(2595)^+$
and $\Lambda_c(2625)^+$, respectively. The full width of the $\Lambda_c(2595)^+$ is fixed to the 
PDG value of 3.6 MeV/$c^2$, and the mass resolution for each peak is fixed to 2.0~MeV/$c^2$, as 
determined from simulation. The fitted signal yields are $9.7\pm3.5$ and $9.3\pm3.2$ for the
$\Lambda_c(2595)^+$ and $\Lambda_c(2625)^+$, respectively. Before the trigger, we find signal
yields of $10.6\pm3.8$ for $\Lambda_c(2595)^+$ and $15.7\pm4.1$ for $\Lambda_c(2625)^+$, 
corresponding to statistical significances of 4.3 and 6.6 standard deviations.
Thus we have evidence for $\Lb\to\Lambda_c(2595)^+\pi^-$ and observation of $\Lb\to\Lambda_c(2625)^+\pi^-$.
The systematic uncertainties do not change this conclusion. These decays have
also been reported by CDF~\cite{cdflam3pi}, but are not yet published.
The fitted $\Delta M_{\pi\pi}$ values of $(306.7\pm1.1)$~MeV/$c^2$ and $(341.7\pm0.6)$~MeV/$c^2$,
for the $\Lambda_c(2625)^+$ and $\Lambda_c(2625)^+$, respectively, are consistent with the known
mass differences~\cite{pdg} for these excited states. 

We also observe the decays $\Lb\to\Sc^{0,++}\pi^{\mp}\pi^-$, with $\Sc^0\to\Lc\pi^-$ or $\Sc^{++}\to\Lc\pi^+$.
The $\Delta M_{\pi}$ distributions are shown in Fig.~\ref{fig:Lb2LcA1_ExcCharm}(b-d) for
both $\Sc^0$ and $\Sc^{++}$ candidates, $\Sc^0$ candidates only (c), and
(d) $\Sc^{++}$ candidates only.
The data are fit to the sum of a Breit-Wigner shape convolved with a Gaussian
resolution function and a smooth threshold function. The full width is fixed to 2.2~MeV/$c^2$~\cite{pdg} in all cases, and the
$\Delta M_{\pi}$ resolution is fixed to 1~MeV/$c^2$ based on simulation. The combined $\Sc^0$ and $\Sc^{++}$ signal
has a statistical significance of 6.0 standard deviations. The $\Sc^0$ and $\Sc^{++}$ signals have statistical significances 
of 4.9 and 3.5, respectively. These decays have also been seen by CDF~\cite{cdflam3pi}.

Table~\ref{tab:excCharm_yields} summarizes the yields for the various excited charm states 
for both the full data sample and after the trigger selection as well as the yields in the 
normalizing modes (after trigger selection.)

\renewcommand{\arraystretch}{1.2}
\begin{table*}[ht]
\begin{center}
\caption{Summary of yields for the signal and normalization modes. Below
$D_1$ and $D_2^*$ refer to the $D_1(2420)$ and $D_2^*(2460)$ mesons, respectively.}
\begin{tabular}{lccc}
\hline\hline
Decay         &       \multicolumn{2}{c}{$\xc^*\pi(\pi)$ Signal Yields} & $\xc\pi^-\pi^+\pi^-$ \\
                                   &       All  & Trig. Sel     & Trig. Sel  \\
\hline
$\Bzb\to D_1^+\pi^-,~D_1^+\to D^+\pi^-\pi^+$              & $41\pm8$     & $33\pm7$    &  $1741\pm55$ \\
$B^-\to D_1^0\pi^-,~D_1^0\to D^0\pi^-\pi^+$                    & $165\pm17$   & $126\pm14$  &  $1386\pm51$ \\
~~$B^-\to D_1^0\pi^-,~D_1^0\to D^{*+}\pi^-$ & $111\pm14$   & $75\pm12$   &  $1386\pm51$ \\
~~$B^-\to D_1^0\pi^-,~D_1^0\to D^0\pi^-\pi^+,~{\rm non-}D^{*}$           & $57\pm10$    & $52\pm9$   &  $1386\pm51$ \\
$B^-\to D_2^{*0}\pi^-,~D_2^{*0}\to D^0\pi^-\pi^+$                    & $66\pm15$    & $49\pm12$   &  $1386\pm51$ \\
~~$B^-\to D_2^{*0}\pi^-,~D_2^{*0}\to D^{*+}\pi^-$ & $46\pm12$    & $34\pm10$   &  $1386\pm51$ \\
~~$B^-\to D_2^{*0}\pi^-,~D_2^{*0}\to D^0\pi^-\pi^+,~{\rm non-}D^{*}$           & $23\pm9$     & $18\pm8$   &  $1386\pm51$ \\
$\Lb\to\Lambda_c(2595)^+\pi^-$           &       $10.6\pm3.8$  &       $9.7\pm3.5$ &  $312\pm23$ \\
$\Lb\to\Lambda_c(2625)^+\pi^-$           &       $15.7\pm4.1$  &       $9.3\pm3.2$ &  $312\pm23$ \\
$\Lb\to\Sc^{0,++}\pi^{\mp}\pi^-$   &      $29.3\pm7.0$  &       $24.9\pm6.2$ &  $312\pm23$ \\
$\Lb\to\Sc^0\pi^-\pi^+$       &       $19.6\pm5.7$  &       $16.2\pm5.0$ &  $312\pm23$ \\
$\Lb\to\Sc^{++}\pi^-\pi^-$    &       $10.1\pm4.0$  &       $9.3\pm3.7$ &  $312\pm23$ \\
\hline\hline
\end{tabular}
\label{tab:excCharm_yields}
\end{center}
\end{table*}
\renewcommand{\arraystretch}{1}

The branching ratios for these modes are computed using:

\begin{equation}
{\br(\xb\to\xc^*\pi(\pi))\times\br(\xc^*\to\xc\pi(\pi))\over \br(\xbtoxcpipipi)} = 
{N_{\rm signal} \over N_{\rm norm}}
\left( \eff_{\rm sel}^{\rm rel}\times \eff_{\rm trig|sel}^{\rm rel}\right)^{-1}
\label{eq:partialbf}
\end{equation}

\noindent where $\xc^*$ refers to one of the observed excited charm states,
$N_{\rm signal}$ and $N_{\rm norm}$
are the number of reconstructed decays in the signal and normalization modes after the trigger requirement,
$\eff_{\rm sel}^{\rm rel}$ is the reconstruction and selection efficiency relative to the
normalization mode, and $\eff_{\rm trig|sel}^{\rm rel}$ is the
relative trigger efficiency. All efficiencies are given for the mass region $0.8~{\rm GeV}/c^2<M(\pi^-\pi^+\pi^-)<3$~GeV/$c^2$. 

The relative reconstruction, selection and trigger efficiencies, shown in Table~\ref{tab:exc_charm},
are evaluated using MC simulations. The $D_1(2420)^0$ and $D_2^*(2460)^0$
are each assumed to decay 70\% through $D^{*+}\pi^-\to D^0\pi^+\pi^-$ and 30\% non-resonant $D^0\pi^+\pi^-$. 
The $D_1(2420)^+$ is taken to be 100\% non-resonant $D^+\pi^-\pi^+$. The $\Lambda_c(2595)^+$ decay is simulated 
as 36\% $\Sc^0\pi^+$, 36\% $\Sc^{++}\pi^-$ and 28\% non-resonant $\Lc\pi^-\pi^+$. 
The $\Lambda_c(2625)^+$ decay is assumed to be 100\% non-resonant $\Lc\pi^-\pi^+$. The $\Sc(2544)$ baryons are
simulated non-resonant in phase space.

The relative efficiencies agree qualitatively with our expectations
based on the kinematics and proximity to threshold for these excited charm states. 
The differences in the relative efficiency
between the pairs of excited charm states for a given $b$-hadron species are negligible
compared to the uncertainty from our limited MC event sample, and we use the average relative 
efficiency for each pair of decays.

\renewcommand{\arraystretch}{1.2}
\begin{table*}[ht]
\begin{center}
\caption{Summary of the relative reconstruction and selection efficiencies ($\eff_{\rm sel}^{\rm rel}$)
and trigger efficiencies ($\eff_{\rm trig|sel}^{\rm rel}$)
for the excited charm hadron intermediate states with respect to the inclusive $\xc\pi^-\pi^+\pi^-$ final 
states. Below $D_1$ and $D_2^*$ refer to $D_1(2420)$ and $D_2^*(2460)$, respectively. The uncertainties shown
are statistical only.}
\begin{tabular}{lccc}
\hline\hline
Decay &          $\eff_{\rm sel}^{\rm rel}$ &  $\eff_{\rm trig|sel}^{\rm rel}$ & $\eff_{\rm total}^{\rm rel}$ \\
                     & (\%) & (\%) & (\%)  \\
\hline
$\Bzb\to D_1^+\pi^-$   &  $0.83\pm0.06$ & $1.05\pm0.09$ & $0.87\pm0.10$ \\ [1ex]
$B^-\to (D_1^0,~D_2^{*0})\pi^-$ & $0.70\pm0.04$ & $1.24\pm0.07$ & $0.86\pm0.07$ \\
~$B^-\to (D_1^0,~D_2^{*0})\pi^- ({\rm via}~D^*)$ & $0.66\pm0.05$ & $1.29\pm0.08$ & $0.84\pm0.08$ \\
~$B^-\to (D_1^0,~D_2^{*0})\pi^- ({\rm non-}D^*)$ & $0.78\pm0.06$ & $1.15\pm0.10$ & $0.91\pm0.11$ \\ [1ex]
$\Lb\to(\Lambda_c(2595),~\Lambda_c(2625)^+)\pi^-$  & $0.52\pm0.03$ & $1.30\pm0.07$ & $0.67\pm0.06$ \\ [1ex]
$\Lb\to\Sc^{0,++}\pi\pi,~\Sc^{0,++}\to\Lc\pi^{\mp}$   & $0.67\pm0.05$ & $1.10\pm0.13$ & $0.75\pm0.10$ \\
\hline\hline
\end{tabular}
\label{tab:exc_charm}
\end{center}
\end{table*}
\renewcommand{\arraystretch}{1.0}

The dominant sources of systematic uncertainty are the limited MC sample sizes and the fit model. 
Starting with the $\bzb$, the uncertainty due to limited MC statistics is 11\%.
For the fit model, the largest source of uncertainty 
is from a possible $D_2^*(2460)^+\pi^-,~D_2^*(2460)^+\to D^+\pi^-\pi^+$ contribution. If this contribution is 
included in the fit using a Breit-Wigner shape with mean and width taken from the PDG~\cite{pdg}, the returned signal 
yield is $0^{+7}_{-0}$. If we assume isospin symmetry, and constrain this fraction (relative to $D_1(2420)$) 
to be $(40\pm11)\%$, the ratio
found for the $B^-$ decay, the fitted $\bzb\to D_1(2420)^+\pi^-,~D_1(2420)^+\to D^+\pi^-\pi^+$ signal yield is $26\pm6$ events.
We take this as a one-sided uncertainty of $^{+0\%}_{-21\%}$. Sensitivity to the background shape is estimated by
using a second order polynomial for the background (3\%). 
The $\bzb$ mass sidebands, which have a $D_1(2420)^+$ fitted yield of $2^{+3}_{-2}$ events
from which we conservatively assign as a one-sided systematic uncertainty of $^{+0\%}_{-6\%}$. 
For the signal decays, 4\% of events have $M(\pi^-\pi^+\pi^-)>3$~GeV/$c^2$, whereas for the $D_1(2420)^+$,
we find a negligible fraction fail this requirement. We therefore apply a correction of $0.96\pm0.02$,
where we have taken 50\% uncertainty on the correction as the systematic error.
The systematic uncertainty on the yield in the $\btodpipipi$ normalizing mode is 3\%. We thus arrive at a total 
systematic error on the $\bzb$ branching fraction ratio of $^{+12}_{-25}\%$.

For the $B^-$, we have a similar set of uncertainties. They are as follows: MC sample size (8\%), 
background model (1\%, 2\%), $D_1(2420)^0$ width (2\%, 4\%), 
$D_2^*(2460)^0$ width (1\%, 3\%), where the two uncertainties are for the ($D_1(2420)^0$, $D_2^*(2460)^0$) 
intermediate states. We have not accounted for interference, and have assumed it is negligible 
compared to other uncertainties. A factor of $0.98\pm0.01$ is applied to correct for the fraction
of events with $M(\pi^-\pi^+\pi^-)>3$~GeV/$c^2$. Including a 3\% uncertainty on the
$\btodzeropipipi$ yield, we find total systematic errors of 9\% and 10\% for the $D_1(2420)^0$ and 
$D_2^*(2460)^0$ intermediate states, respectively. For the $D^*$ sub-decays, the total systematic uncertainties are 
10\% and 11\% for $B^-\to D_1(2420)^0\pi^-,~D_1(2420)^0\to D^{*+}\pi^-$ and $B^-\to D_2^*(2460)^0\pi^-,~D_2^*(2460)^0\to D^{*+}\pi^-$,
respectively. For final states not through $D^*$, we find a total systematic uncertainty of 13\% for
both intermediate states. In all cases, the dominant systematic uncertainty is the limited number of 
MC events.

For the $\Lb$ branching fraction ratios, we attribute uncertainty to limited MC sample sizes (8\%),
the $\Lc(2595)$ width ($^{+9\%}_{-5\%}$), $\LbtoLcpipipi$ signal yield (3\%), and apply a
correction of $0.96\pm0.02$ for the ratio of yields with $M(\pi^-\pi^+\pi^-)>3$~GeV/$c^2$.
In total, the systematic uncertainties on the $\Lc(2595)^+$ and $\Lambda_c(2625)^+$ partial branching 
fractions are $_{-10\%}^{+13\%}$ and $\pm10\%$, respectively.

For the $\Sc^{0,++}$ intermediate states, the systematic uncertainties include 14\% from finite MC statistics,
and 4\% from the $\Sigma_c^{0,++}$ width. For the $\Sc^{0,++}$ simulation, 10\% of decays have $M(\pi^-\pi^+\pi^-)>3$~GeV/$c^2$,
compared to 4\% for the normalizing mode. We therefore apply a correction of $1.06\pm0.03$ to the ratio of 
branching fractions. All other uncertainties are negligible in comparison.
We thus arrive at a total systematic uncertainty of 16\%.

The final partial branching fractions are
\begin{align*}
{\br(\Bzb\to D_1^-\pi^+,~D_1^-\to D^+\pi^-\pi^+) \over \btodpipipi} &= (2.1\pm0.5_{-0.5}^{+0.3})\% \nonumber \\
{\br(B^-\to D_1^0\pi^+,~D_1^0\to D^0\pi^-\pi^+) \over \btodzeropipipi} &= (10.3\pm1.5\pm0.9)\% \nonumber \\
{\br(B^-\to D_1^0\pi^+,~D_1^0\to D^{*+}\pi^-) \over \btodzeropipipi} &= (9.3\pm1.6\pm0.9)\% \nonumber \\
{\br(B^-\to D_1^0\pi^+,~D_1^0\to D^0\pi^-\pi^+)_{{\rm non-}D^{*}} \over \btodzeropipipi} &= (4.0\pm0.7\pm0.5)\% \nonumber \\
{\br(B^-\to D_2^{*0}\pi^+,~D_2^{*0}\to D^0\pi^-\pi^+) \over \btodzeropipipi} &= (4.0\pm1.0\pm0.4)\% \nonumber \\
{\br(B^-\to D_2^{*0}\pi^+,~D_2^{*0}\to D^{*+}\pi^-) \over \btodzeropipipi} &= (3.9\pm1.2\pm0.4)\% \nonumber \\
{\br(B^-\to D_2^{*0}\pi^+,~D_2^{*0}\to D^0\pi^-\pi^+)_{{\rm non-}D^{*}} \over \btodzeropipipi} &= (1.4\pm0.6\pm0.2)\%  \nonumber \\
& <3.0\% {\rm~at~90\%~C.L.} \nonumber \\ 
{\br(\Lb\to\Lambda_c(2595)^+\pi^+,~\Lambda_c(2595)^+\to\Lc\pi^-\pi^+) \over \LbtoLcpipipi} &= (4.4\pm1.7_{-0.4}^{+0.6})\% \nonumber\\
{\br(\Lb\to\Lambda_c(2625)^+\pi^+,~\Lambda_c(2625)^+\to\Lc\pi^-\pi^+) \over \LbtoLcpipipi} &= (4.3\pm1.5\pm0.4)\% \nonumber \\
{\br(\Lb\to\Sc^{0,++}\pi^{\mp}\pi^-,~\Sc^{0,++}\to\Lc\pi^{\mp}) \over \LbtoLcpipipi} &= (11.4\pm3.1\pm1.8)\% \nonumber \\
{\br(\Lb\to\Sc^{0}\pi^+\pi^-,~\Sc^0\to\Lc\pi^-) \over \LbtoLcpipipi} &= (7.4\pm2.4\pm1.2)\% \nonumber \\
{\br(\Lb\to\Sc^{++}\pi^-\pi^-,~\Sc^{++}\to\Lc\pi^+) \over \LbtoLcpipipi} &= (4.2\pm1.8\pm0.7)\%, \nonumber \\
\end{align*}

\noindent where the first uncertainties are statistical and the second are systematic. 
For the modes with $D^{*+}$, we include a factor $\br(D^{*+}\to D^0\pi^+)=(0.677\pm0.005)$~\cite{pdg}
to account for unobserved $D^{*+}$ decays.  
The first four and the sixth of these decays have been previously measured by Belle~\cite{belleb0tod2420pi}
with comparable precision. To compare our results to those absolute branching fractions, 
we multiply them by the relative $\bzb$ [$B^-$] branching fractions in Eq.~\ref{eq:cfbf}, and then in turn
by $\br(\btodpi)=(2.68\pm0.13)\times10^{-3}$ [$\br(\btodzeropi) = (4.84\pm0.15)\times10^{-3}$.]
The resulting absolute branching fractions are

\begin{align*}
\br(\Bzb\to D_1(2420)^-\pi^+,~D_1(2420)^-\to D^+\pi^-\pi^+) = (1.3\pm0.3_{-0.3}^{+0.2})\times10^{-4} \nonumber \\
\br(B^-\to D_1(2420)^0\pi^+,~D_1(2420)^0\to D^0\pi^-\pi^+) = (6.3\pm0.9\pm0.9)\times10^{-4} \nonumber \\
\br(B^-\to D_1(2420)^0\pi^+,~D_1(2420)^0\to D^{*+}\pi^-) = (5.8\pm1.0\pm0.9)\times10^{-4} \nonumber \\
\br(B^-\to D_1(2420)^0\pi^+,~D_1(2420)^0\to D^0\pi^+\pi^-)_{{\rm non-}D^*} = (2.5\pm0.4\pm0.4)\times10^{-4} \nonumber \\
\br(B^-\to D_2^*(2460)^0\pi^+,~D_2^*(2460)^0\to D^{*+}\pi^-) = (2.5\pm0.7\pm0.4)\times10^{-4} \nonumber \\
\end{align*}

\noindent where the uncertainties are statistical and total systematic, respectively.  
The corresponding values obtained by Belle are:
$(0.89^{+0.23}_{-0.35})\times10^{-4}$, $(6.5^{+1.1}_{-1.2})\times10^{-4}$, $(6.8\pm1.5)\times10^{-4}$,
$(1.9^{+0.5}_{-0.6})\times10^{-4}$, and  $(1.8\pm0.5)\times10^{-4}$~\cite{pdg,belleb0tod2420pi}. 
Our results are consistent with, and of comparable precision to, those measurements.

Preliminary results on the $\Lb\to\Lc(2595)^+\pi^-$, $\Lb\to\Lc(2625)^+\pi^-$ and $\Lb\to\Sc^{0,++}\pi^{\mp}\pi^-$  
decays have been reported by CDF~\cite{cdflam3pi}. Our values are consistent with these (unpublished) results.

\section{Summary}

In summary, we have measured the branching fractions for $\xbtoxcpipipi$ decays relative to $\xbtoxcpi$.
The ratio of branching fractions are measured to be

\begin{align*}
{\br(\btodpipipi)\over\br(\btodpi)} = 2.38\pm0.11\pm0.21 \nonumber \\
{\br(\btodzeropipipi)\over\br(\btodzeropi)} = 1.27\pm0.06\pm0.11 \nonumber \\
{\br(\bstodspipipi)\over\br(\bstodspi)} = 2.01\pm0.37\pm0.20 \nonumber \\
{\br(\LbtoLcpipipi)\over\br(\LbtoLcpi)} = 1.43\pm0.16\pm0.13. \nonumber \\
\end{align*}

\noindent At low 3$\pi$ mass, these decays appear to be dominated by the $a_1(1260)$
resonance. We have also measured several partial decay rates through excited charm states.
The yields of $\xbtoxcpipipi$ relative to $\xbtoxcpi$ are in the range of 20$-$40\%.
If the relative rates in the Cabibbo-suppressed decays, such as $\bstodskpipi$ and $\btodzerokpipi$
relative to $\bstodsk$ and $\btodzerok$, respectively, are comparable, they could be useful
for measuring the weak phase $\gamma$. 

\section*{Acknowledgments}

\noindent We express our gratitude to our colleagues in the CERN accelerator
departments for the excellent performance of the LHC. We thank the
technical and administrative staff at CERN and at the LHCb institutes,
and acknowledge support from the National Agencies: CAPES, CNPq,
FAPERJ and FINEP (Brazil); CERN; NSFC (China); CNRS/IN2P3 (France);
BMBF, DFG, HGF and MPG (Germany); SFI (Ireland); INFN (Italy); FOM and
NWO (Netherlands); SCSR (Poland); ANCS (Romania); MinES of Russia and
Rosatom (Russia); MICINN, XuntaGal and GENCAT (Spain); SNSF and SER
(Switzerland); NAS Ukraine (Ukraine); STFC (United Kingdom); NSF
(USA). We also acknowledge the support received from the ERC under FP7
and the Region Auvergne.


\clearpage

\end{document}